\documentclass[a4paper]{article}
\usepackage[a4paper, margin=1in]{geometry}

\usepackage[colorlinks=true, citecolor=green, linkcolor=black, urlcolor=blue]{hyperref}
\usepackage[font=footnotesize]{caption}
\usepackage{authblk}
\usepackage{graphicx, wrapfig, rotating}
\usepackage{lineno}
\usepackage{enumitem}
\usepackage{booktabs}
\usepackage{amsmath}
\usepackage{amssymb}
\usepackage{stmaryrd} 
\usepackage{mathtools}
\usepackage{float}
\usepackage{bm}
\usepackage{xr}
\usepackage{tikz}
\usetikzlibrary{matrix,decorations.pathreplacing, calc, positioning,fit}
\usepackage[style=numeric-comp, url=false, doi=true, sorting=none, isbn = false, maxcitenames=2, maxbibnames=30, giveninits=true, date=year]{biblatex}

\renewbibmacro{in:}{%
  \ifentrytype{article}{}{\printtext{\bibstring{in}\intitlepunct}}}
  
\renewbibmacro{in:}{, In:}

\DeclareNameAlias{author}{last-first}

\DeclareFieldFormat[article, inbook, incollection,inproceedings,patent,thesis,unpublished]{title}{#1} 

\DeclareFieldFormat[article]{pages}{#1}

\providecommand{\keywords}[1]{{Keywords:} #1}

\begin{document}

\pagenumbering{arabic}
\date{}
\title{A comprehensive comparison of total-order estimators for global sensitivity analysis}
\author[1,2]{Arnald Puy\thanks{Corresponding author}}
\author[3]{William Becker}
\author[4]{Samuele Lo Piano}
\author[5]{Andrea Saltelli}

\affil[1]{\footnotesize{\textit{Department of Ecology and Evolutionary Biology, M31 Guyot Hall, Princeton University, New Jersey 08544, USA. E-Mail: \href{mailto:apuy@princeton.edu}{apuy@princeton.edu}}}}

\affil[2]{\footnotesize{\textit{Centre for the Study of the Sciences and the Humanities (SVT), University of Bergen, Parkveien 9, PB 7805, 5020 Bergen, Norway.}}}

\affil[3]{\footnotesize{\textit{European Commission, Joint Research Centre, Via Enrico Fermi, 2749, 21027 Ispra VA, Italy}}}

\affil[4]{\footnotesize{\textit{School of the Built Environment, JJ Thompson Building, University of Reading, Whiteknights Campus, Reading, RG6 6AF, United Kingdom}}}

\affil[5]{\footnotesize{\textit{Open Evidence Research, Universitat Oberta de Catalunya (UOC), Barcelona, Spain.}}}

\maketitle


\begin{abstract}
Sensitivity analysis helps identify which model inputs convey the most uncertainty to the model output. One of the most authoritative measures in global sensitivity analysis is the Sobol' total-order index, which can be computed with several different estimators. Although previous comparisons exist, it is hard to know which estimator performs best since the results are contingent on the benchmark setting defined by the analyst (the sampling method, the distribution of the model inputs, the number of model runs, the test function or model and its dimensionality, the weight of higher order effects or the performance measure selected). Here we compare several total-order estimators in an eight-dimension hypercube where these benchmark parameters are treated as random parameters. This arrangement significantly relaxes the dependency of the results on the benchmark design. We observe that the most accurate estimators are Razavi and Gupta's, Jansen's or Janon/Monod's for factor prioritization, and Jansen's, Janon/Monod's or Azzini and Rosati's for approaching the ``true'' total-order indices. The rest lag considerably behind. Our work helps analysts navigate the myriad of total-order formulae by reducing the uncertainty in the selection of the most appropriate estimator.
\end{abstract}

\keywords{Uncertainty analysis; sensitivity analysis; modeling; Sobol' indices; variance decomposition, benchmarking analysis}
 
 \newpage
\section{Introduction}  
\label{sec:introduction}
Sensitivity analysis, i.e. the assessment of how much uncertainty in a given model output is conveyed by each model input, is a fundamental step to judge the quality of model-based inferences \cite{Saltelli2008, Jakeman2006, Eker2018}. Among the many sensitivity indices available, variance-based indices are widely regarded as the gold standard because they are model-free (no assumptions are made about the model), global (they account for interactions between the model inputs) and easy to interpret \cite{Saltelli2002b, Iooss2015, Becker2014}. Given a model of the form $y=f(\bm{x})$, $\bm{x}=(x_1, x_2, ...,x_i,..., x_k)\in \mathbb{R}^k$, where $y$ is a scalar output and $x_1,...,x_k$ are the $k$ independent model inputs, the variance of $y$ is decomposed into conditional terms as
\begin{equation}
V(y)=\sum_{i=1}^{k}V_i+\sum_{i}\sum_{i<j}V_{ij}+...+V_{1,2,...,k} \\,
\label{eq:decomposition}
\end{equation}
where 
\begin{equation}
\begin{aligned}
V_i = V_{x_{i}}\big[E_{\bm{x}_{\sim i}}(y | x_i)\big] \hspace{4mm} 
V_{ij} &= V_{x_{i}, x_{j}}\big[E_{\bm{x}_{\sim i, j}}(y | x_i, x_j)\big] \hspace{4mm}\\
& - V_{x_{i}}\big[E_{\bm{x}_{\sim i}}(y | x_i)\big] \\
& - V_{x_{j}}\big[E_{\bm{x}_{\sim j}}(y | x_j)\big]
\end{aligned}
\label{eq:Ex_i}
\end{equation}
and so on up to the $k$-th order.  The notation $\bm{x}_{\sim i}$ means \emph{all-but-$x_i$}. By dividing each term in Equation~\ref{eq:decomposition} by the unconditional model output variance $V(y)$, we obtain the first-order indices for single inputs ($S_i$), pairs of inputs ($S_{ij}$), and for all higher-order terms. First-order indices thus provide the proportion of $V(y)$ caused by each term and are widely used to rank model inputs according to their contribution to the model output uncertainty, a setting known as \emph{factor prioritization} \parencite{Saltelli2008}.

\textcite{Homma1996a} also proposed the calculation of the total-order index $T_i$, which measures the first-order effect of a model input jointly with its interactions up to the $k$-th order:

\begin{equation}
T_i=1 - \frac{V_{\bm{x}_{\sim i}}\big[E_{x_i}(y | \bm{x}_{\sim i})\big]}{V(y)} = \frac{E_{\bm{x}_{\sim i}}\big[V_{x_{i}}(y | \bm{x}_{\sim i})\big]}{V(y)} \,.
\label{eq:Ti}
\end{equation}

When $T_i \approx 0$, it can be concluded that $x_i$ has a negligible contribution to $V(y)$. For this reason, total-order indices have been applied to distinguish influential from non-influential model inputs and reduce the dimensionality of the uncertain space, a setting known as \emph{factor-fixing} \parencite{Saltelli2008}. 

The most direct computation of $T_i$ is via Monte Carlo (MC) estimation because it does not impose any assumption on the functional form of the response function, unlike metamodeling approaches \parencite{LeGratiet2017, Saltelli1999}. The Fourier Amplitude Sensitivity Test (FAST) may also be used to calculate $T_i$, which involves transforming input variables into periodic functions of a single frequency variable, sampling the model and analysing the sensitivity of input variables using Fourier analysis in the frequency domain \cite{cukier1973study, cukier1978nonlinear}. While an innovative approach, FAST is sensitive to the characteristic frequencies assigned to input variables, and is not a very intuitive method -  for these reasons it has mostly been superseded by Monte Carlo approaches, or by metamodels when computational expense is a serious issue. In this work we focus on the former.

MC methods require generating a $(N, 2k)$ base sample matrix with either random or quasi-random numbers (e.g. Latin Hypercube Sampling, Sobol' quasi-random numbers \cite{Sobol1967, Sobol1976}), where each row is a sampling point and each column a model input. The first $k$ columns are allocated to an $\bm{A}$ matrix and the remaining $k$ columns to a $\bm{B}$ matrix, which are known as the ``base sample matrices''. Any point in either $\bm{A}$ or $\bm{B}$ can be indicated as $x_{vi}$, where $v$ and $i$ respectively index the row (from 1 to $N$) and the column (from 1 to $k$). Then, $k$ additional $\bm{A}_{B}^{(i)}$ ($\bm{B}_{A}^{(i)}$) matrices are created, where all columns come from $\bm{A}$ ($\bm{B}$) except the $i$-th column, which comes from $\bm{B}$ ($\bm{A}$). The numerator in Equation~\ref{eq:Ti} is finally estimated using the model evaluations obtained from the $\bm{A}$ ($\bm{B}$) and  $\bm{A}_{B}^{(i)}$ ($\bm{B}_{A}^{(i)}$) matrices. Some estimators may also use a third or $\bm{X}$ base sample matrices (i.e. $\bm{A}, \bm{B}, \bm{C}, \hdots, \bm{X}$), although the use of more than three matrices has been recently proven inefficient by \textcite{LoPiano2021}.

\subsection{Total-order estimators and uncertainties in the benchmark settings}
The search for efficient and robust total-order estimators is an active field of research \cite{Homma1996a, Jansen1999, Saltelli2008, Janon2014, Glen2012, Azzini2020, Monod2006a, Razavi2016a}. Although some works have compared their asymptotic properties (i.e. \cite{Janon2014}), most studies have promoted empirical comparisons where different estimators are benchmarked against known test functions and specific sample sizes. However valuable these empirical studies may be, \textcite{Becker2020} observed that their results are very much conditional on the choice of model, its dimensionality and the selected number of model runs. It is hard to say from previous studies whether an estimator outperforming another truly reflects its higher accuracy or simply its better performance under the narrow statistical design of the study. Below we extend the list of factors which \textcite{Becker2020} regards as influential in a given benchmarking exercise and discuss how they affect the relative performance of sensitive estimators.

\begin{itemize}
\item \textit{The sampling method:} The creation of the base sample matrices can be done using Monte-Carlo (MC) or quasi Monte-Carlo (QMC) methods \cite{Sobol1967, Sobol1976}. Compared to MC, QMC allows to more effectively map the input space as it leaves smaller unexplored volumes (Fig.~S1). However, \textcite{Kucherenko2011} observed that MC methods might help obtain more accurate sensitivity indices when the model under examination has important high-order terms. Both MC and QMC have been used when benchmarking sensitivity indices \cite{Jansen1999, Saltelli2010a}.

\item \textit{The form of the test function:} some of the most commonly used functions in SA  are the \textcite{Ishigami1990}'s, the Sobol' G and its variants \parencite{Sobol1998, Saltelli2010a}, the \textcite{Bratley1988a}'s or the set of functions presented in \textcite{Kucherenko2011} \parencite{Janon2014, Azzini2020, Saltelli2010a, LoPiano2021}. Despite being analytically tractable, these functions capture only one possible interval of model behaviour, and the effects of nonlinearities and nonadditivities is typically unknown in real models. This \emph{black-box} nature of models has become more of a concern in the last decades due to the increase in computational power and code complexity (which prevents the analyst from intuitively grasping the model's behaviour \parencite{Borgonovo2016}), and to the higher demand for model transparency \parencite{Eker2018, Saltelli2019b, Saltelli2020a}. This renders the functional form of the model similar to a random variable \parencite{Becker2020}, something not accounted for by previous works \parencite{Janon2014, Azzini2020, Saltelli2010a, LoPiano2021}.

\item \textit{The function dimensionality:} many studies focus on low-dimensional problems, either by using test functions that only require a few model inputs (e.g. the Ishigami function, where $k=3$), or by using test functions with a flexible dimensionality, but setting $k$ at a small value of e.g. $k\leq8$ (\textcite{Sobol1998}'s G or \textcite{Bratley1988a} functions). This approach trades computational manageability for comprehensiveness: by neglecting higher dimensions, it is difficult to tell which estimator might work best in models with tens or hundreds of parameters. Examples of such models can be readily found in the Earth and Environmental Sciences domain \parencite{Sheikholeslami2019a}, including the Soil and Water Assessment Tool (SWAT) model, where $k=50$ \parencite{Sarrazin2016}, or the Mod\'{e}lisation Environmentale-Surface et Hydrologie (MESH) model, where $k=111$ \parencite{Haghnegahdar2017}.

\item \textit{The distribution of the model inputs}: the large majority of benchmarking exercises assume uniformly-distributed inputs $p(\bm{x})\in U(0,1)^k$ \parencite{Janon2014, Azzini2019, Saltelli2010a, LoPiano2021}. However, there is evidence that the accuracy of $T_i$ estimators might be sensitive to the underlying model input distributions, to the point of overturning the model input ranks \parencite{Shin2013, Paleari2016}. Furthermore, in uncertainty analysis --  e.g. in decision theory, the analysts may use distributions with peaks for the most likely values derived, for instance, from an experts elicitation stage.

\item \textit{The number of model runs:} sensitivity test functions are generally not computationally expensive and can be run without much concern for computational time. This is frequently not the case for real models, whose high dimensionality and complexity might set a constraint on the total number of model runs available. Under such restrictions, the performance of the estimators of the total-order index depends on their efficiency (how accurate they are given the budget of runs that can be allocated to each model input). There are no specific guidelines as to which total-order estimator might work best under these circumstances \parencite{Becker2020}.
\item \textit{The performance measure selected:} typically, a sensitivity estimator has been considered to outperform the rest if, on average, it displays a smaller mean absolute error (MAE), computed as 

\begin{equation}
\text{MAE}=\frac{1}{p}\sum_{v=1}^{p} \left ( \frac{\sum_{i=1}^{k} | T_i - \hat{T}_i|}{k} \right )\,,
\label{eq:MAE}
\end{equation}
where $p$ is the number of replicas of the sample matrix,  and $T_i$ and $\hat{T}_i$ the analytical and the estimated total-order index of the $i$-th input. The  MAE is appropriate when the aim is to assess which estimator better approaches the true total-order indices, because it averages the error for both influential and non-influential indices. However, the analyst might be more interested in using the estimated indices $\bm{\hat{T}}=\{\hat{T}_1,\hat{T}_2,...,\hat{T}_i,...,\hat{T}_k\}$ to accurately rank parameters or screen influential from non-influential model inputs \parencite{Saltelli2008}. In such context, the MAE may be best substituted or complemented with a measure of rank concordance between the vectors  $\bm{r}$ and $\bm{\hat{r}}$, which reflect the ranks in  $\bm{T}$ and $\bm{\hat{T}}$ respectively, such as the Spearman's $\rho$ or the Kendall's $W$ coefficient \parencite{Spearman1904, Kendall1939, Becker2020}. It can also be the case that disagreements on the exact ranking of low-ranked parameters may have no practical importance because the interest lies in the correct identification of top ranks only \parencite{Sheikholeslami2019a}. \textcite{Savage1956} scores or other measures that emphasize this top-down correlation are then a more suitable choice. 
\end{itemize}

Here we benchmark the performance of eight different MC-based formulae available to estimate $T_i$ (Table~1). While the list is not exhaustive, they reflect the research conducted on $T_i$ over the last 20 years: from the classic estimators of \textcite{Homma1996a, Jansen1999, Saltelli2008} up to the new contributions by \textcite{Janon2014}, \textcite{Glen2012}, \textcite{Azzini2019} and \textcite{Razavi2016a, Razavi2016b}. In order to reduce the influence of the benchmarking design in the assessment of the estimators' accuracy, we treat the sampling method $\tau$, the underlying model input distribution $\phi$, the number of model runs $N_t$, the test function $\varepsilon$, its dimensionality and degree of non-additivity ($k,k_2,k_3$) and the performance measure $\delta$ as random parameters. This better reflects the diversity of models and sensitivity settings available to the analyst. By relaxing the dependency of the results on these benchmark parameters\footnote{We refer to the set of benchmarking assumptions as \emph{benchmarking parameters} or \emph{parameters}. This is intended to distinguish them from the inputs of each test function generated by the metafunction, which we refer to as inputs.}, we define an unprecedentedly large setting where all formulae can prove their accuracy. We therefore extend \textcite{Becker2020}'s approach by testing a wider set of Monte Carlo estimators, by exploring a wider range of benchmarking assumptions and by performing a formal SA on these assumptions. The aim is therefore to provide a much more global comparison of available MC estimators than is available in the existing literature, and investigate how the benchmarking parameters may affect the relative performance of estimators. Such information can help point to estimators that are not only efficient on a particular case study, but efficient and robust to a wide range of practical situations.

\begingroup
\renewcommand{\arraystretch}{1.7}
\begin{table}[!ht]
\centering
\caption{Formulae to compute $T_i$. $f_0$ and $V(y)$ are estimated according to the original papers. For estimators 2 and 5, $f_0=\frac{1}{N}\sum_{v=1}^{N}f(\bm{A})_v$. For estimators 1, 2 and 5, $V(y)=\frac{1}{N}\sum_{v=1}^{N} \left [ f(\bm{A})_v-f_0 \right]^2$ \cites[Eq.~4.16]{Saltelli2008}[Eqs.~15, 20]{Homma1996a}. For estimator 3, $f_0 = \frac{1}{N} \sum_{v=1}^{N}\frac{f(\bm{A})_v + f(\bm{A}_{B} ^{(i)})_v}{2} $ and $V(y)=\frac{1}{N} \sum_{v=1}^{N} \frac{f(\bm{A})_v^2 + f(\bm{A}_{B} ^{(i)})_v^2}{2} - f_0^2$ \cite[Eq.~15]{Janon2014}. In estimator 4, $\langle f(\bm{A})_v \rangle$ is the mean of $f(\bm{A})_v$. We use a simplified version of the Glen and Isaacs estimator because spurious correlations are zero by design. As for estimator 7, we refer to it as pseudo-Owen given its use of a $\bm{C}$ matrix and its identification with \textcite{Owen2013} in \textcite{Iooss2020}, where we retrieve the formula from. $V(y)$ in Estimator 7 is computed as in Estimator 3 following \textcite{Iooss2020}, whereas $V(y)$ in Estimator 8 is computed as in Estimator 1.}
\label{tab:Ti_estimators}
\begin{tabular}{llp{3.2cm}}
\toprule
Nº & Estimator & Author \\
\midrule
1 & $\frac{\frac{1}{2N} \sum_{v=1}^{N } \left [ f(\bm{A})_v - f(\bm{A}_{B} ^{(i)})_v \right ] ^ 2}{V(y)}$ & \textcite{Jansen1999} \\
2 &  $\frac{V(y) - \frac{1}{N} \sum_{v = 1}^{N} f(\bm{A})_v f(\bm{A}_{B} ^{(i)})_v + f_0^2}{V(y)}$ & \textcite{Homma1996a} \\
3 & $1 - \frac{\frac{1}{N} \sum_{v=1}^{N}f(\bm{A})_v f(\bm{A}_{B} ^{(i)})_v - f_0^2}{V(y)}$ & \textcite{Janon2014} \newline \textcite{Monod2006a}  \\
4 &  $ 1 - \left [\frac{1}{N-1}\sum_{v=1}^{N} \frac{\left [ f(\bm{A})_v - \left \langle f(\bm{A})_v\right \rangle \right ] \left[ f(\bm{A}_{B} ^{(i)})_v - \left \langle f(\bm{A}_{B} ^{(i)})_v\right \rangle \right ]}{\sqrt{V\left [f(\bm{A})_v\right ] V\left [f(\bm{A}_{B} ^{(i)})_v \right]}} \right ]
$ &  \textcite{Glen2012} \\
5 & $1 - \frac{\frac{1}{N} \sum_{v=1}^{N} f(\bm{B})_v f(\bm{B}^{(i)}_A)_v - f_0^2}{V(y)}$ & \textcite{Saltelli2008} \\
6 & $\frac{\sum_{v = 1}^{N} [ f(\bm{B})_v - f(\bm{B}^{(i)}_A)_v ] ^ 2 +   [ f(\bm{A})_v - f(\bm{A}^{(i)}_B)_v  ] ^ 2}{\sum_{v = 1} ^ {N} [ f(\bm{A})_v - f(\bm{B}) _v  ] ^ 2 +  [ f(\bm{B}^{(i)}_A)_v - f(\bm{A}^{(i)}_B)_v  ] ^ 2}$ & \textcite{Azzini2019, Azzini2020} \\
7 & $\frac{V(y) -  \left [ \frac{1}{N} \sum_{v=1} ^ {N} \left \{\left [ f(\bm{B})_v - f(\bm{C}_{B} ^{(i)})_v \right ] \left [ f(\bm{B}_{A} ^{(i)})_v - f(\bm{A})_v \right ] \right \} \right ]}{V(y)}$ &  pseudo-Owen \\
8 & $\frac{E_{x^*_{\sim_{i}}}\left [ \gamma_{x^*{_\sim i}}(h_i)\right] + E_{x^*{_\sim i}} \left [ C_{x^*{_\sim i}}(h_i) \right ] }{V(y)}$  & \textcite{Razavi2016a, Razavi2016b} (see SM). \\
\bottomrule
\end{tabular}
\end{table}
\endgroup

\section{Assessment of the uncertainties in the benchmarking parameters}
\label{sec:materials}

In this section we formulate the benchmarking parameters as random variables and assess how the performance of estimators is dependent on them by performing a sensitivity analysis. In essence this is a \emph{sensitivity analysis of sensitivity analyses} \cite{Puy2020}, and a natural extension of a similar uncertainty analysis in a recent work by \textcite{Becker2020}. The use of global sensitivity analysis tools to better understand the properties of estimators can give insights into how estimators behave in different scenarios that are not available through analytical approaches.

\subsection{The setting}

The variability in the benchmark settings ($\tau, N_t, k, k_2, k_3, \phi, \epsilon, \delta$) is described by probability distributions (Table~\ref{tab:parameters}). We assign uniform distributions (discrete or continuous) to each parameter. In particular, we choose $\tau\sim\mathcal{DU}(1, 2)$ to check how the performance of $T_i$ estimators is conditioned by the use of Monte-Carlo ($\tau=1$) or Quasi Monte-Carlo ($\tau=2$) methods in the creation of the base sample matrices. For $\tau=2$ we use the Sobol' sequence scrambled according to \textcite{Owen1995} to avoid repeated coordinates at the beginning of the sequence. The total number of model runs and inputs is respectively described as $N_t\sim\mathcal{DU}(10, 1000)$ and $k\sim\mathcal{DU}(3,100)$  to explore the performance of the estimators in a wide range of $N_t,k$ combinations. Given the sampling constraints set by the estimators' reliance on either a $\bm{B}$, $\bm{B}_{A} ^{(i)}$, $\bm{A}_{B} ^{(i)}$ or $\bm{C}_{B} ^{(i)}$ matrices (Table~\ref{tab:Ti_estimators}), we modify the space defined by ($N_t, k$) to a non-rectangular domain (we provide more information on this adjustment in Section~\ref{sec:algorithm}). 

\begin{table}[ht]
\centering
\caption{Summary of the parameters and their distributions. $\mathcal{DU}$ stands for discrete uniform.}
\label{tab:parameters}
\begin{tabular}{llc}
\toprule
Parameter & Description & Distribution \\
\midrule
$\tau$ & Sampling method & $\mathcal{DU}(1, 2)$ \\
$N_t$ & Total number of model runs & $\mathcal{DU}(10,1000)$\\
$k$ & Number of model inputs & $\mathcal{DU}(3,100)$ \\
$\phi$ & Probability distribution of the model inputs & $\mathcal{DU}(1, 8)$ \\
$\varepsilon$ & Randomness in the test function & $\mathcal{DU}(1, 200)$ \\
$k_2$ & Fraction of pairwise interactions & $\mathcal{U}(0.3, 0.5)$ \\
$k_3$ & Fraction of three-wise interactions & $\mathcal{U}(0.1, 0.3)$ \\
$\delta$ & Selection of the performance measure & $\mathcal{DU}(1, 2)$ \\
\bottomrule
\end{tabular}
\end{table}

\begin{figure}[ht]
\centering
\includegraphics[keepaspectratio]{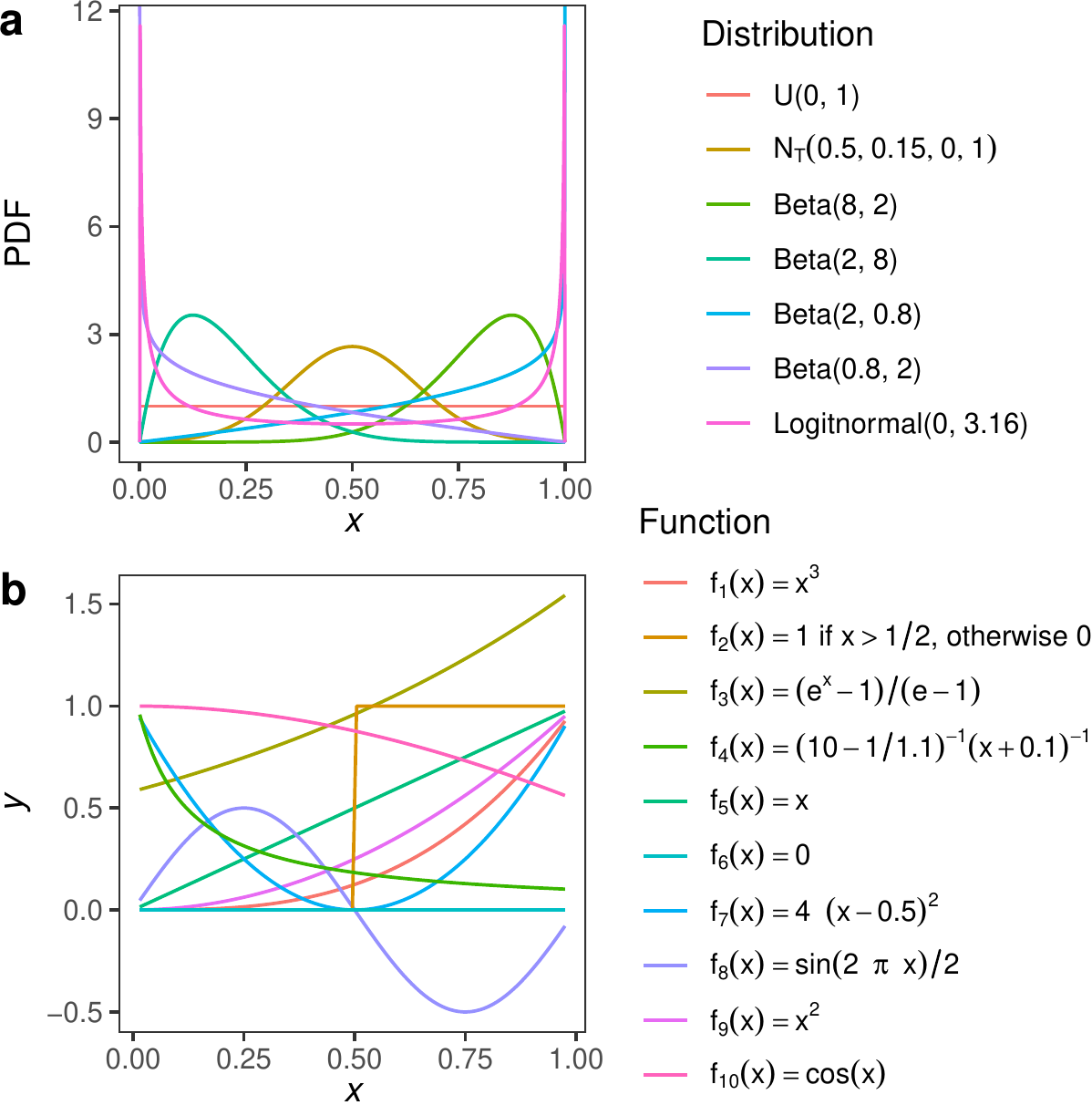}
\caption{The metafunction approach. a) Probability distributions incuded in $\phi$. $N_T$ stands for truncated normal distribution. b) Univariate functions included in the metafunction ($f_1(x)=$ cubic, $f_2(x)=$ discontinuous, $f_3(x)=$ exponential, $f_4(x)=$ inverse, $f_5(x)=$ linear, $f_6(x)=$ no effect, $f_7(x)=$ non-monotonic, $f_8(x)=$ periodic, $f_9(x)=$ quadratic, $f_{10}(x)=$ trigonometric).}
\label{fig:metafunction}
\end{figure}

For $\phi$ we set $\phi\sim\mathcal{DU}(1,8)$ to ensure an adequate representation of the most common shapes in the $(0,1)^k$ domain. Besides the normal distribution truncated at $(0,1)$ and the uniform distribution, we also take into account four beta distributions parametrized with distinct $\alpha$ and $\beta$ values and a logitnormal distribution (Fig.~\ref{fig:metafunction}a). The aim is to check the response of the estimators under a wide range of probability distributions, including U-shaped distributions and distributions with different degrees of skewness.

We link each distribution in Fig.~\ref{fig:metafunction}a to an integer value from 1 to 7. For instance, if $\phi=1$, the joint probability distribution of the model inputs is described as $p(x_1,\hdots,x_k)=\mathcal{U}(0,1)^k$. If $\phi=8$, we create a vector $\bm{\phi}=\{\phi_1,\phi_2,...,\phi_i,...,\phi_k\}$ by randomly sampling the seven distributions in Fig.~\ref{fig:metafunction}a, and use the $i$-th distribution in the vector to describe the uncertainty of the $i$-th input. This last case examines the behavior of the estimators when several distributions are used to characterize the uncertainty in the model input space.

\subsubsection{The test function}

The parameter $\varepsilon$ operationalizes the randomness in the form and execution of the test function. Our test function is an extended version of \textcite{Becker2020}'s metafunction, which randomly combines $p$ univariate functions in a multivariate function of dimension $k$. Here we consider the 10 univariate functions listed in Fig.~\ref{fig:metafunction}b, which represent common responses observed in physical systems and in classic SA test functions (see \textcite{Becker2020} for a discussion on this point). We note that an alternative approach would be to construct orthogonal basis functions which could allow analytical evaluation of true sensitivity indices for each generated function; however, this extension is left for future work.

We construct the test function as follows:

\begin{enumerate}
\item Let us consider a sample matrix such as 

\begin{equation}
\bm{M}= 
\begin{bmatrix}
x_{11} &  x_{12} & \cdots  & x_{1i} & \cdots & x_{1k} \\
x_{21} &  x_{22} & \cdots  & x_{2i} & \cdots &  x_{2k} \\
\vdots & \vdots & \ddots & \vdots &  \ddots & \vdots \\
x_{v1} &  x_{v2} & \cdots  & x_{vi} & \cdots &  x_{vk} \\
\vdots & \vdots & \ddots & \vdots &  \ddots & \vdots \\
x_{N1} &  x_{N2} & \cdots  & x_{Ni} & \cdots &  x_{Nk} \\
\end{bmatrix}
\end{equation}

where every point $\bm{x}_v=x_{v1}, x_{v2}, \hdots, x_{vk}$ represents a given combination of values for the $k$ inputs and $x_i$ is a model input whose distribution is defined by $\phi$.

\item Let $\bm{u}=\{u_1,u_2,...,u_k\}$ be a $k$-length vector formed by randomly sampling with replacement the ten functions in Fig.~\ref{fig:metafunction}b. The $i$-th function in $\bm{u}$ is then applied to the $i$-th model input: for instance, if $k=4$ and $\bm{u}=\{u_3,u_4,u_8, u_1\}$, then $f_3(x_1)=\frac{e^{x_1}-1}{e-1}$, $f_4(x_2)=(10-\frac{1}{1.1})^{-1}(x_2 + 0.1)^{-1}$, $f_8(x_3)=\frac{\sin(2\pi x_3)}{2}$, and $f_1(x_4)=x_4^3$. The elements in $\bm{u}$ thus represent the first-order effects of each model input.

\item Let $\bm{V}$ be a $(n, 2)$ matrix, for $n=\frac{k!}{2!(k-2)!}$, the number of pairwise combinations between the $k$ inputs of the model. Each row in $\bm{V}$ thus specifies an interaction between two columns in $\bm{M}$. In the case of $k=4$ and the same elements in $\bm{u}$ as defined in the previous example, 

\begin{equation}
\bm{V}= 
\begin{bmatrix}
1 & 2\\
1 & 3 \\
1 & 4 \\
2 & 3 \\
2 & 4 \\
3 & 4 \\
\end{bmatrix}
\end{equation}

e.g., the first row promotes $f_3(x_1) \cdot f_4(x_2)$, the second row $f_3(x_1) \cdot  f_8(x_3)$, and so on until the $n$-th row. In order to follow the \emph{sparsity of effects principle} (most variations in a given model output should be explained by low-order interactions \parencite{Box2005}), the metafunction activates only a fraction of these effects: it randomly samples $\llceil k_2n \rrceil $ rows from $\bm{V}$, and computes the corresponding interactions in $\bm{M}$. $\llceil k_2n \rrceil $ is thus the number of pairwise interactions present in the function. We make $k_2$ an uncertain parameter described as $k_2\sim\mathcal{U}(0.3, 0.5)$ in order to randomly activate only between 30\% and 50\% of the available second-order effects in $\bm{M}$.

\item Same as before, but for third-order effects: let $\bm{W}$ be a ($m, 3$) matrix, for $m=\frac{k!}{3!(k-3)!}$, the number of three-wise combinations between the $k$ inputs in $\bm{M}$. For $k=4$ and $\bm{u}$ as before, 

\begin{equation}
\bm{W}= 
\begin{bmatrix}
1 & 2 & 3\\
1 & 2 & 4 \\
1 & 3 & 4 \\
2 & 3 & 4 \\
\end{bmatrix}
\end{equation}
e.g. the first row leads to $f_3(x_1)\cdot f_4(x_2) \cdot f_8(x_3)$, and so on until the $m$-th row. The metafunction then randomly samples $\llceil k_3m \rrceil$ rows from $\bm{W}$ and computes the corresponding interactions in $\bm{M}$. $\llceil k_3m \rrceil $ is therefore the number of three-wise interaction terms in the function. We also make $k_3$ an uncertain parameter described as $k_3\sim\mathcal{U}(0.1,0.3)$ to activate only between 10\% and 30\% of all third-order effects in $\bm{M}$. Note that $k_2>k_3$ because third-order effects tend to be less dominant than two-order effects (Table~\ref{tab:parameters}).

\item Three vectors of coefficients ($\bm{\alpha}, \bm{\beta}, \bm{\gamma}$) of length $k$, $n$ and $m$ are defined to represent the weights of the first, second and third-order effects respectively. These coefficients are generated by sampling from a mixture of two normal distributions $\Psi=0.3\mathcal{N}(0, 5) + 0.7\mathcal{N}(0, 0.5)$. This coerces the metafunction into replicating the \textcite{Pareto1906} principle (around 80\% of the effects are due to 20\% of the parameters), found to widely apply in SA \parencite{Box1986, Saltelli2008}.

\item The metafunction can thus be formalized as \begin{equation}
\begin{aligned}
y = & \sum_{i=1}^{k}\alpha_i f^{u_i}\phi_i(x_i) \\
& + \sum_{i=1}^{\llceil k_2n \rrceil}\beta_i f^{u_{V_{i,1}}} \phi_i(x_{V_{i,1}}) f^{u_{V_{i,2}}} \phi_i(x_{V_{i,2}}) \\
& + \sum_{i=1}^{\llceil k_3m \rrceil}\gamma_i f^{u_{W_{i,1}}} \phi_i(x_{W_{i,1}}) f^{u_{W_{i,2}}} \phi_i(x_{W_{i,2}}) f^{u_{W_{i,3}}} \phi_i(x_{W_{i,3}})\,.
\end{aligned}
\label{eq:metafunction}
\end{equation}

Note that there is randomness in the sampling of $\bm{\phi}$, the univariate functions in $\bm{u}$ and the coefficients in $(\bm{\alpha}, \bm{\beta}, \bm{\gamma})$. The parameter $\varepsilon$ assesses the influence of this randomness by fixing the starting point of the pseudo-random number sequence used for sampling the parameters just mentioned. We use $\varepsilon\sim\mathcal{U}(1,200)$ to ensure that the same seed does not overlap with the same value of $N_t$, $k$ or any other parameter, an issue that might introduce determinism in a process that should be stochastic. In Figs.~S2--S3we show the type of $T_i$ indices generated by this metafunction.

\end{enumerate}

Finally, we describe the parameter $\delta$ as $\delta\sim\mathcal{DU}(1,2)$. If $\delta=1$, we compute the Kendall $\tau$-b correlation coefficient between $\bm{\hat{r}}$ and $\bm{r}$, the estimated and the ``true'' ranks calculated from $\bm{\hat{T}}$ and $\bm{T}$ respectively. This aims at evaluating how well the estimators in Table~\ref{tab:Ti_estimators} rank all model inputs. If $\delta=2$, we compute the Pearson correlation between $\bm{r}$ and $\bm{\hat{r}}$ after transforming the ranks to Savage scores \cite{Savage1956}. This setting examines the performance of the estimators when the analyst is interested in ranking only the most important model inputs. Savage scores are given as
\begin{equation}
Sa_i=\sum_{j=i}^{k}\frac{1}{j}\,,
\label{eq:savage_scores}
\end{equation}
where $j$ is the rank assigned to the $j$th element of a vector of length $k$. If $x_1>x_2>x_3$, the Savage scores would then be $Sa_1=1+\frac{1}{2}+\frac{1}{3}$, $Sa_2=\frac{1}{2}+\frac{1}{3}$, and $Sa_3=\frac{1}{3}$. The parameter $\delta$ thus assesses the accuracy of the estimators in properly ranking the model inputs; in other words, when they are used in a factor prioritization setting \cite{Saltelli2008}.

In order to examine also how accurate the estimators are in approaching the ``true'' indices, we run an extra round of simulations with the MAE as the only performance measure,  which we compute as 
\begin{equation}
\text{MAE}= \frac{\sum_{i=1}^{k} | T_i - \hat{T}_i|}{k}\,.
\label{eq:MAE_no}
\end{equation}
Note that, unlike Equation~\ref{eq:MAE}, Equation~\ref{eq:MAE_no} does not make use of replicas. This is because the effect of the sampling is averaged out in our design by simultaneously varying all parameters in many different simulations.

\subsection{The execution of the algorithm}
\label{sec:algorithm}

We examine how sensitive the performance of total-order estimators is to the uncertainty in the benchmark parameters $\tau, N_t, k, k_2, k_3, \phi, \epsilon, \delta$ by means of a global SA. We create an $\bm{A}$, $\bm{B}$ and $k-1$ $\bm{A}^{(i)}_B$ matrices, each of dimension $(2^{11}, k)$, using Sobol' quasi-random numbers. In these matrices each column is a benchmark parameter described with the probability distributions of Table~\ref{tab:parameters} and each row is a simulation with a specific combination of $\tau, N_t, k,\hdots$ values. Note that we use $k-1$ $\bm{A}^{(i)}_B$ matrices because we group $N_t$ and $k$ and treat them like a single benchmark parameter given their correlation (see below).

Our algorithm runs rowwise over the $\bm{A}$, $\bm{B}$ and $k-1$ $\bm{A}^{(i)}_B$ matrices, for $v=1,2,\hdots,18,432$ rows. In the $v$-th row it does the following:

\begin{enumerate}
\item It creates five $(N_{t_v},k_v)$ matrices using the sampling method defined by $\tau_v$. The need for these five sub-matrices responds to the five specific sampling designs requested by the estimators of our study (Table~\ref{tab:Ti_estimators}). We use these matrices to compute the vector of estimated indices $\bm{\hat{T}}_i$ for each estimator:

\begin{enumerate}
\item An $\bm{A}$ matrix and $k_v$ $\bm{A}_{B}^{(i)}$ matrices, each of size $(N_v, k_v)$, $N_v=\llceil \frac{N_{t{_v}}}{k_v+1} \rrceil$ (Estimators 1--4 in Table~\ref{tab:Ti_estimators}).

\item An $\bm{A}$, $\bm{B}$ and $k_v$ $\bm{A}_{B}^{(i)}$ matrices, each of size $(N_v, k_v)$, $N_v=\llceil \frac{N_{t{_v}}}{k_v+ 2} \rrceil$ (Estimator 5 in Table~\ref{tab:Ti_estimators}).

\item  An $\bm{A}$, $\bm{B}$ and $k_v$ $\bm{A}_{B}^{(i)}$ and $\bm{B}_{A}^{(i)}$ matrices, each of size $(N_v, k_v)$, $N_v=\llceil \frac{N_{t_{v}}}{2k_v+2} \rrceil$ (Estimator 6 in Table~\ref{tab:Ti_estimators}).

\item  An $\bm{A}$, $\bm{B}$ and $k_v$ $\bm{B}_{A}^{(i)}$ and $\bm{C}_{B}^{(i)}$ matrices, each of size $(N_v, k_v)$, $N_v=\llceil \frac{N_{t{_v}}}{2k_v+2} \rrceil$  (Estimator 7 in Table~\ref{tab:Ti_estimators}).

\item  A matrix formed by $N_v$ stars, each of size $k_v (\frac{1}{\Delta h} - 1) + 1$. Given that we set $\Delta h$ at 0.2 (see Supplementary Materials), $N_v=\llfloor \frac{N_{t_{v}}}{4k+1} \rrfloor$ (Estimator 8 in Table~\ref{tab:Ti_estimators}).
\end{enumerate}

The different sampling designs and the value for $k_v$ constrains the total number of runs $N_{t_v}$ that can be allocated to each estimator. Furthermore, given the probability distributions selected for $N_t$ and $k$ (Table~\ref{tab:parameters}), specific combinations of ($N_{t_{v}}, k_v$) lead to $N_v\leq1$, which is computationally unfeasible. To minimize these issues we force the comparison between estimators to approximate the same $N_{t_{v}}$ value. Since the sampling design structure of Razavi and Gupta is the most constraining, we use $N_v=\frac{2(4k+1)}{k+1}$ (for estimators 1--4), $N_v=\frac{2(4k+1)}{k+2}$ (for estimator 5) and $N_v=\frac{2(4k+1)}{2k+2}$ (for estimators 6--7) when  $N_v\leq1$ in the case of Razavi and Gupta. This compels all estimators to explore a very similar portion of the ($N_t, k$) space, but $N_t$ and $k$ become correlated, which contradicts the requirement of independent inputs characterizing variance-based sensitivity indices \cite{Saltelli2008}. This is why we treat ($N_t,k$)  as a single benchmark parameter in the SA.

\item It creates a sixth matrix, formed by an $\bm{A}$ and $k_v$ $\bm{A}_{B}^{(i)}$ matrices, each of size $(2^{11}, k_v)$. We use this sub-matrix to compute the vector of ``true'' indices $\bm{T}$, which could not be calculated analytically due to the wide range of possible functional forms created by the metafunction. Following \textcite{Becker2020}, we assume that a fairly accurate approximation to $\bm{T}$ could be achieved with a large Monte Carlo estimation.

\item The distribution of the model inputs in these six sample matrices is defined by $\phi_v$.

\item The metafunction runs over these six matrices simultaneously, with its functional form, degree of active second and third-order effects as set by $\varepsilon_v$, $k_{2_v}$ and $k_{3_v}$ respectively.

\item It computes the estimated sensitivity indices $\bm{\hat{T}}_v$ for each estimator and the ``true'' sensitivity indices $\bm{T}_v$ using the \textcite{Jansen1999} estimator, which is currently best practice in SA.

\item It checks the performance of the estimators. This is done in two ways:

\begin{enumerate}
\item If $\delta=1$, we compute the correlation between $\bm{\hat{r}}_v$ and $\bm{r}_v$ (obtained respectively from $\bm{\hat{T}}_v$ and $\bm{T}_v$)  with Kendall tau, and if $\delta=2$ we compute the correlation between $\bm{\hat{r}}_v$ and $\bm{r}_v$ on Savage scores. The model output in both cases is the correlation coefficient $r$, with higher $r$ values indicating a better performance in properly ranking the model inputs.

\item We compute the MAE between $\bm{\hat{T}}_v$ and $\bm{T}_v$. In this case the model output is the MAE, with lower values indicating a better performance in approaching the ``true'' total-order indices.
\end{enumerate}
\end{enumerate}

\section{Results}

\subsection{Uncertainty analysis}
\begin{figure}[ht]
\centering
\includegraphics[keepaspectratio]{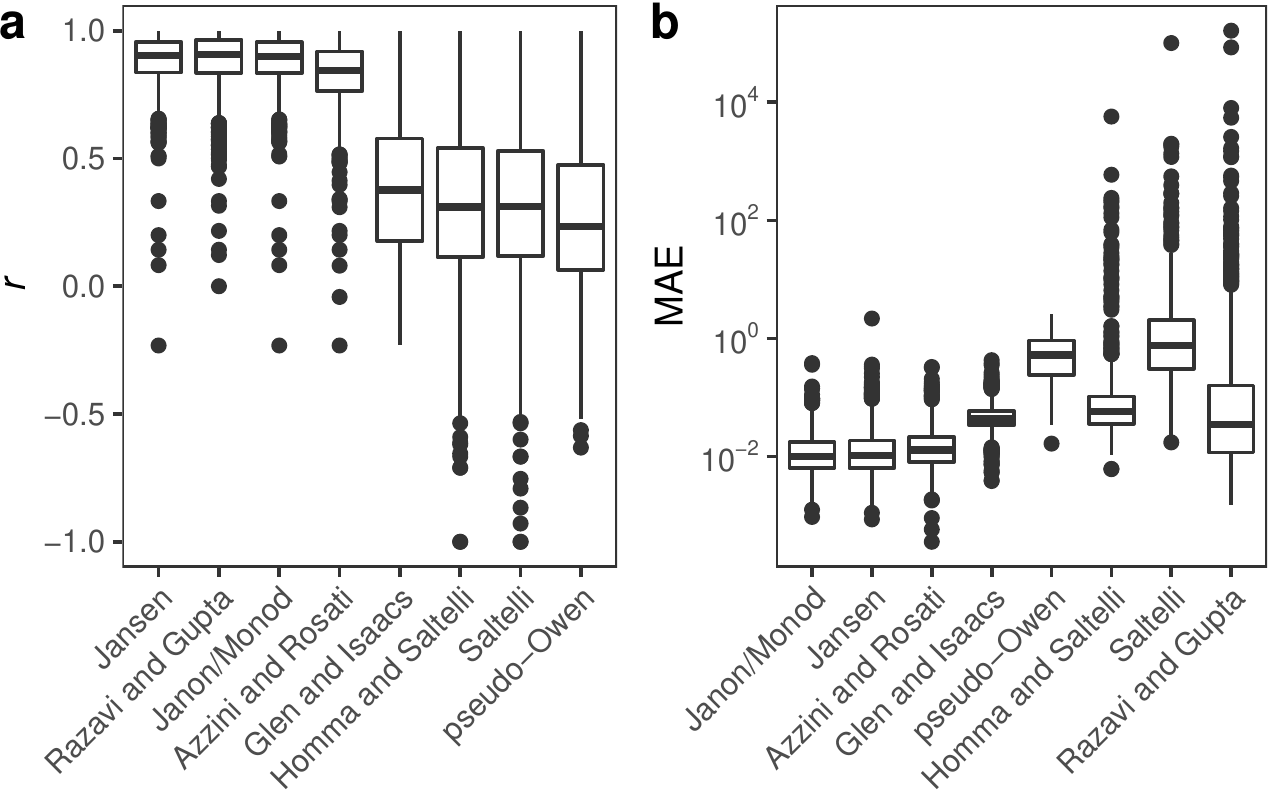}
\caption{Boxplots summarizing the results of the simulations. a) Correlation coefficient between $\bm{\hat{r}}$ and $\bm{r}$, the vector of estimated and ``true'' ranks. b) Mean Absolute Error (MAE).}
\label{fig:boxplots}
\end{figure}

Under a factor prioritization setting (e.g. when the aim is to rank the model inputs in terms of their contribution to the model output variance), the most accurate estimators are Jansen, Razavi and Gupta, Janon/Monod and Azzini and Rosati. The distribution of $r$ values (the correlation between estimated and "true" ranks) when these estimators are used is highly negatively skewed, with median values of $\approx0.9$. Glen and Isaacs, Homma and Saltelli, Saltelli and pseudo-Owen lag behind and display median $r$ values of $\approx0.35$, with pseudo-Owen ranking last ($r\approx0.2$). The range of values obtained with these formulae is much more spread out and include a significant number of negative $r$ values, suggesting that they overturned the true ranks in several simulations (Figs.~\ref{fig:boxplots}a, S4). 

When the goal is to approximate the ``true'' indices, Janon/Monod, Jansen and Azzini and Rosati also offer the best performance. The median MAE obtained with these estimators is generally smaller than Glen and Isaacs' and pseudo-Owen's, and the distribution of MAE values is much more narrower than that obtained with Homma and Saltelli, Saltelli or Razavi and Gupta. These three estimators are the least accurate and produce several MAE values larger than $10^2$ in several simulations (Fig.~\ref{fig:boxplots}b). The volatility of Razavi and Gupta under the MAE is reflected in the numerous outliers produced and sharply contrasts with its very good performance in a factor prioritization setting (Fig.~\ref{fig:boxplots}a).

\begin{figure}[!ht]
\centering
\includegraphics[keepaspectratio]{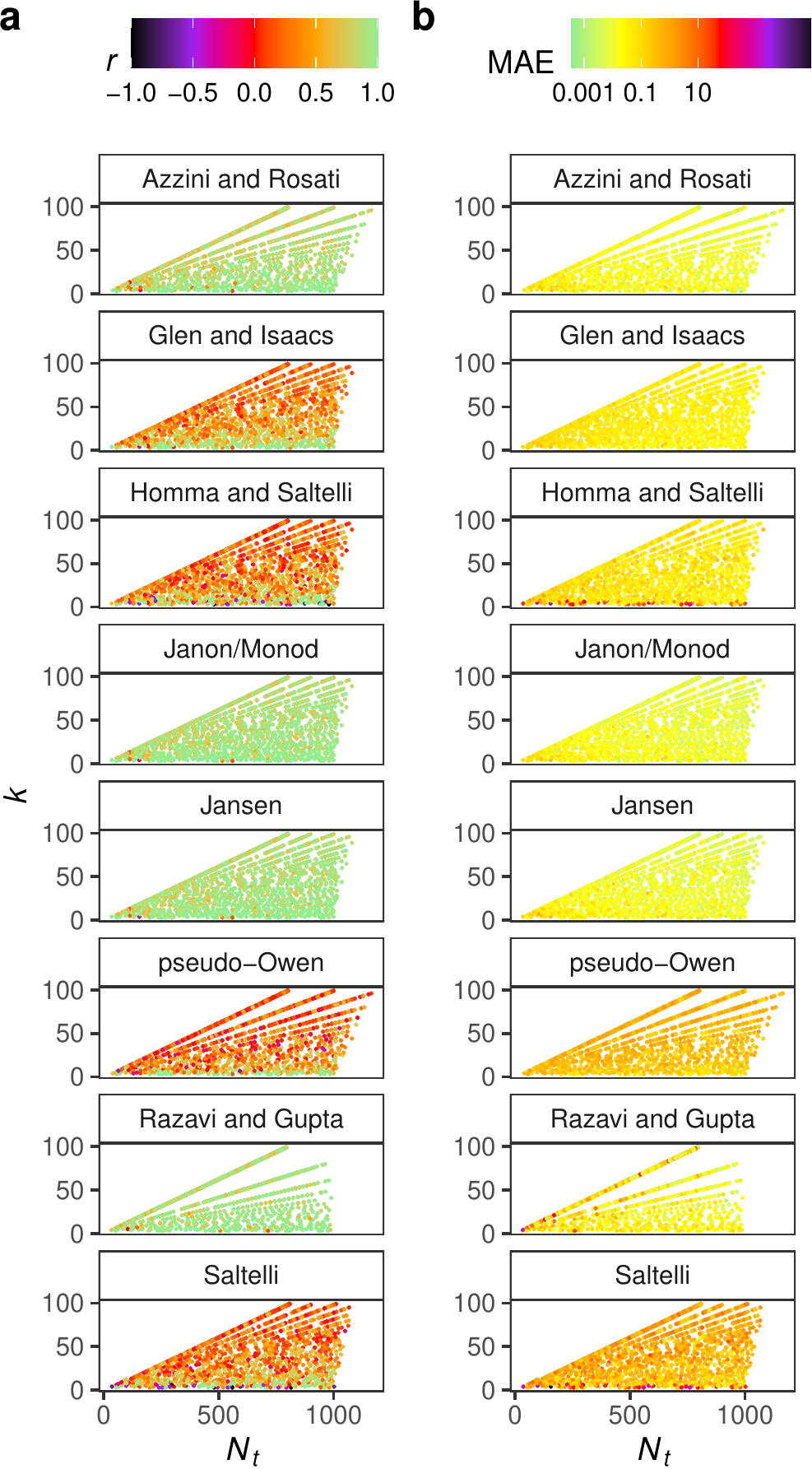}
\caption{Number of runs $N_t$ against the function dimensionality $k$. Each dot is a simulation with a specific combination of the benchmark parameters in Table~\ref{tab:parameters}. The greener (blacker) the color, the better (worse) the performance of the estimator. a) Accuracy of the estimators when the goal is to properly rank the model inputs, e.g. a factor prioritization setting. b) Accuracy of the estimators when the goal is to approach the ``true'' total-order indices.}
\label{fig:scatter_color}
\end{figure}

To obtain a finer insight into the structure of these results, we plot the total number of model runs $N_t$ against the function dimensionality $k$ (Fig.~\ref{fig:scatter_color}). This maps the performance of the estimators in the input space formed by all possible combinations of $N_t$ and $k$ given the specific design constraints of each formulae. Under a factor prioritization setting, almost all estimators perform reasonably well at a very small dimensionality ($k\leq10, r >0.7$), regardless of the total number of model runs available. However, some differences unfold at higher dimensions: Saltelli, Homma and Saltelli, Glen and Isaacs and especially pseudo-Owen swiftly become inaccurate for $k>10$, even with large values for $N_t$. Azzini and Rosati display a very good performance overall except in the upper $N_t,k$ boundary, where most of the orange dots concentrate. The estimators of Jansen, Janon/Monod and Razavi and Gupta rank the model inputs almost flawlessly regardless of the region explored in the $N_t,k$ domain (Fig.~\ref{fig:scatter_color}a).

With regards to the MAE, Janon/Monod, Jansen and Azzini and Rosati maintain their high performance regardless of the $N_t,k$ region explored. The accuracy of Razavi and Gupta, however, drops at the upper-leftmost part of the $N_t,k$ boundary, where most of the largest MAE scores are located ($\mbox{MAE}>10$). In the case of Saltelli and Homma and Saltelli, the largest MAE values concentrate in the region of small $k$ regardless of the total number of model runs, a domain in which they achieved a high performance when the focus was on properly ranking the model inputs.

The presence of a non-negligible proportion of model runs with $r<0$ suggests that some estimators significantly overturned the true ranks  (Figs~\ref{fig:scatter_color}a, S4). To better examine this phenomenon, we re-plot Fig~\ref{fig:scatter_color}b with just the simulations yielding $r<0$ (Fig.~S5). We observe that $r<0$ values not only appear in the region of small $N_t$, a foreseeable miscalculation derived from allocating an insufficient number of model runs to each model input: they also emerge at a relatively large $N_t$ and low $k$ in the case of pseudo-Owen, Saltelli and Homma and Saltelli. The Saltelli estimator actually concentrates in the $k<10$ zone most of the simulations with the lowest negative $r$ values (Fig.~S5). This suggests that rank reversing is not an artifact of our study design as much as a by-product of the volatility of these estimators when stressed by the sources of computational uncertainty listed in Table~\ref{tab:parameters}. Such strain may lead these estimators to produce a significant fraction of negative indices or indices beyond 1, thus effectively promoting $r<0$.

\begin{figure}
\centering
\includegraphics[keepaspectratio]{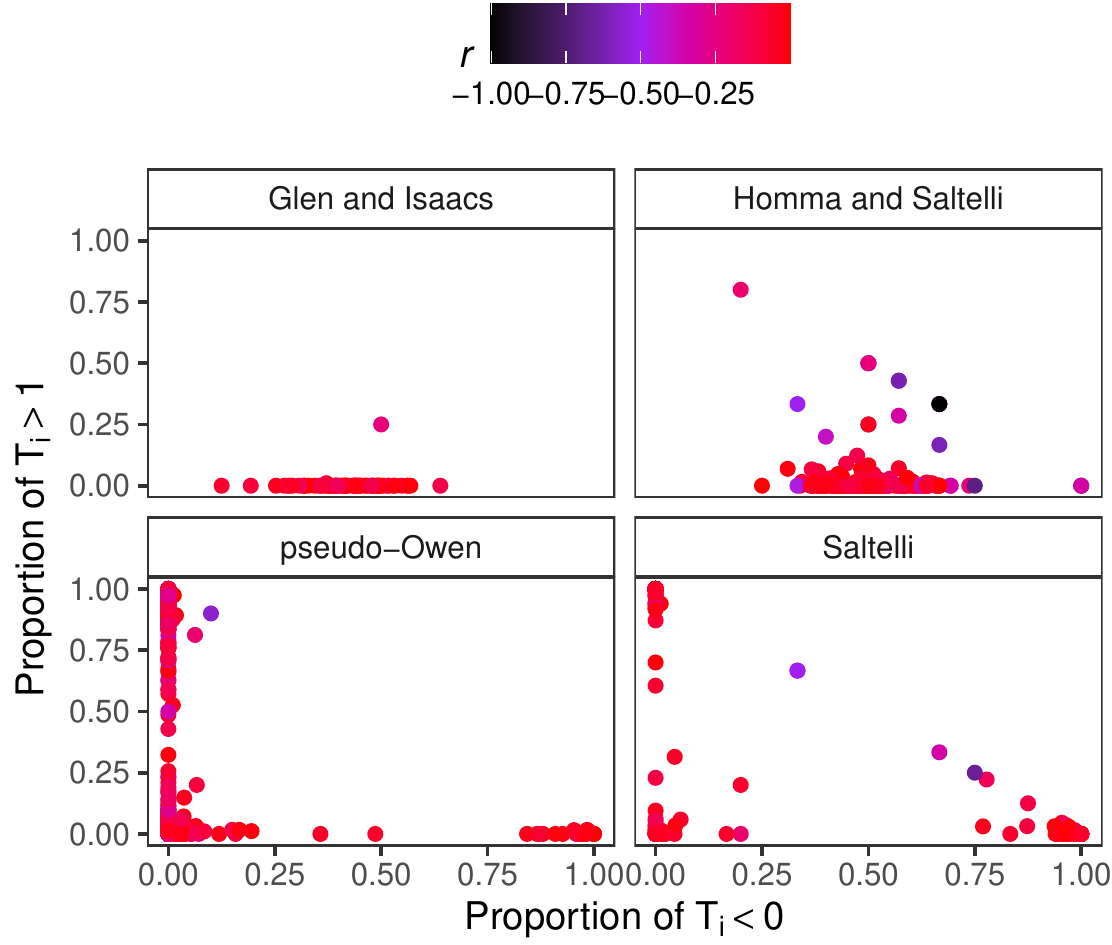}
\caption{Scatterplot of the proportion of $T_i<0$ against the proportion of $T_i>1$ mapped against the model output $r$. Each dot is a simulation. Only simulations with $r<0$ are displayed.}
\label{fig:ti>1}
\end{figure}

We calculate the proportion of $T_i<0$ and $T_i > 1$ in each simulation that yielded $r<0$. In the case of Glen and Isaacs and Homma and Saltelli,  $r<0$ values are caused by the production of a large proportion of $T_i < 0$ (25\%--75\%, the $x$ axis in Fig.~\ref{fig:ti>1}). Pseudo-Owen and Saltelli suffer this bias too and in several simulations they also generate a large proportion of $T_i>1$ (up to 100\% of the model inputs, the $y$ axis in Fig.~\ref{fig:ti>1}). The production of $T_i<0$ and $T_i>1$ is caused by numerical errors and fostered by the values generated at the numerator of Equation~\ref{eq:Ti}: $T_i < 0$ may either derive from $E_{\bm{x}_{\sim i}}\big[V_{x_{i}}(y | \bm{x}_{\sim i})\big] < 0$ (e.g. Homma and Saltelli and pseudo-Owen) or $V_{\bm{x}_{\sim i}}\big[E_{x_i}(y | \bm{x}_{\sim i})\big] > V(y)$ (e.g. Saltelli), whereas $T_i > 1$ from $E_{\bm{x}_{\sim i}}\big[V_{x_{i}}(y | \bm{x}_{\sim i})\big] > V(y)$ (e.g. Homma and Saltelli and pseudo-Owen) or $V_{\bm{x}_{\sim i}}\big[E_{x_i}(y | \bm{x}_{\sim i})\big] < 0$ (e.g. Saltelli).

\begin{figure}[ht]
\centering
\includegraphics[keepaspectratio]{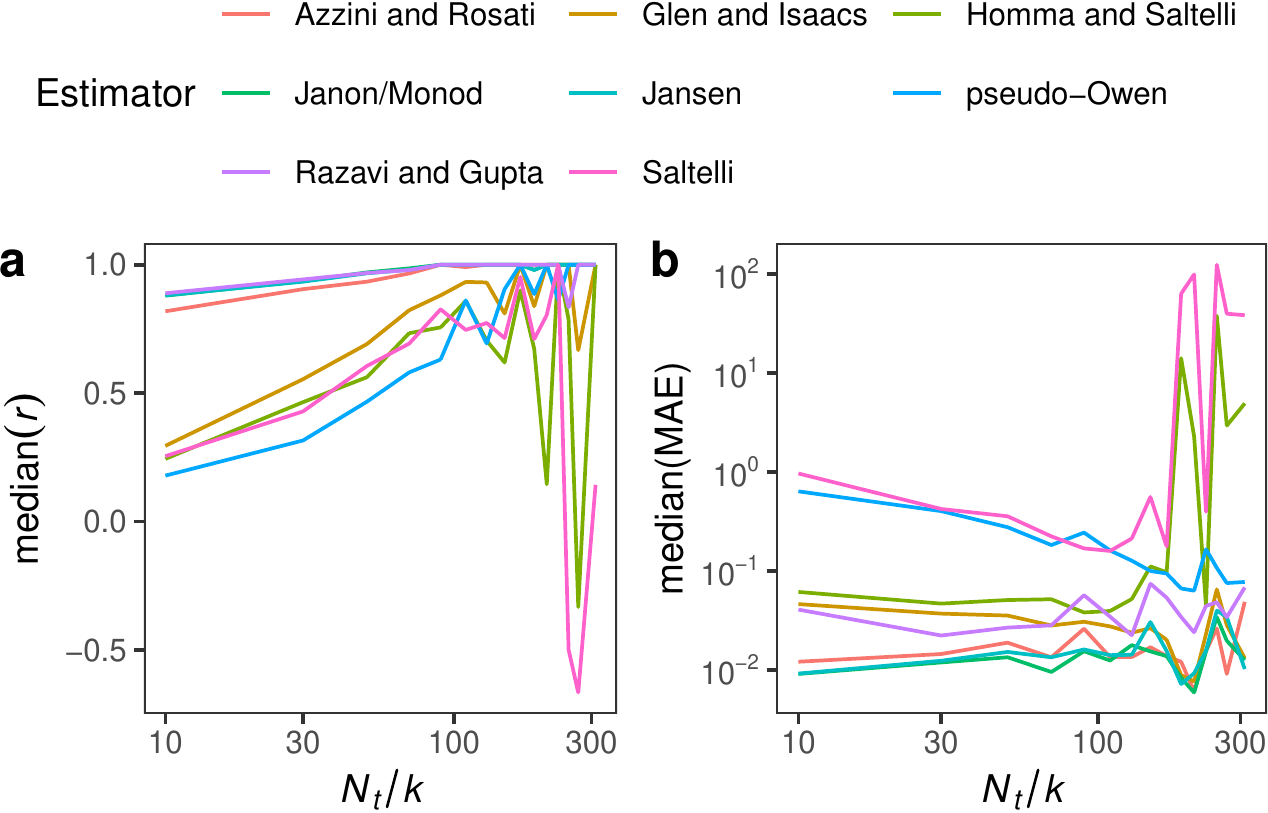}
\caption{Scatterplot of the model output $r$ against the number of model runs allocated per model input $(Nt / k)$. See Fig.~S6  for a visual display of all simulations and Fig.~S7 for an assessment of the number of model runs that each estimator has in each $N_t/k$ compartment.}
\label{fig:medians}
\end{figure}

To better examine the efficiency of the estimators, we summarized their performance as a function of the number of runs available per model input $N_t/k$ \parencite{Becker2020} (Fig.~\ref{fig:medians}, S6). This information is especially relevant to take an educated decision on which estimator to use in a context of a high-dimension, computationally expensive model. Even when the budget of runs per input is low $\left [ (N_t/k) \in [2, 20] \right ]$, Razavi and Gupta, Jansen and Janon/Monod are very good at properly ranking model inputs ($r\approx0.9$), and are followed very close by Azzini and Rosati ($r\approx0.8$). Saltelli, Homma and Saltelli and Glen and Isaacs come after ($r\approx0.3$), with pseudo-Owen scoring last ($r\approx0.2$). When the $N_t/k$ ratio is increased, all estimators improve their ranking accuracy and some quickly reach the asymptote: this is the case of Razavi and Gupta, Janon/Monod and Jansen, whose performance becomes almost flawless from $(N_t/k) \in [40, 60]$ onwards, and of Azzini and Rosati, which reaches its optimum at $(N_t/k) \in [60, 80]$. The accuracy of the other estimators does not seem to fully stabilize within the range of ratios examined. In the case of Homma and Saltelli and Saltelli, their performance oscillates before plummeting at $(N_t/k) \in [200, 210]$, $(N_t/k) \in [240, 260]$ and $(N_t/k) \in [260, 280]$  due to several simulations yielding large $r<0$ values (Fig.~\ref{fig:medians}a).

Janon/Monod and Jansen are also the most efficient estimators when the MAE is the measure of choice, followed closely by Azzini and Rosati, Razavi and Gupta and Glen and Isaacs. Saltelli and Homma and Saltelli gain accuracy at higher $N_t / k$ ratios yet their precision diminishes all the same from $(N_t/k) \in [200, 210]$ onwards (Fig.~\ref{fig:medians}b).

\subsection{Sensitivity analysis}

\begin{figure}[!ht]
\centering
\includegraphics[keepaspectratio]{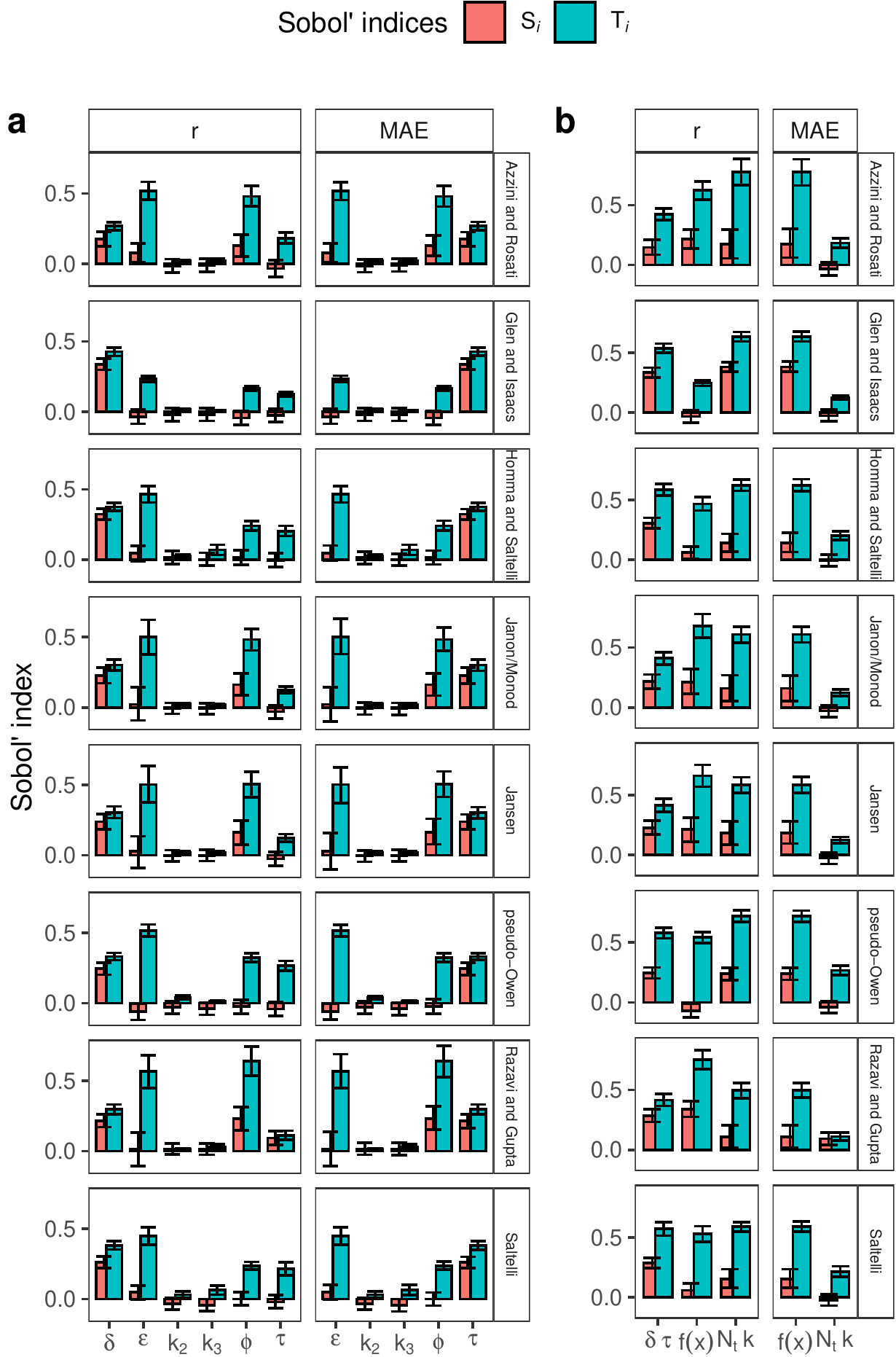}
\caption{Sobol' indices. a) Individual parameters. b) Clusters of parameters. The cluster $f(x)$ includes all parameters that describe the uncertainty in the functional form of the model ($\epsilon, k_2,k_3, \phi$). $N_t$ and $k$ are assessed simultaneously due to their correlation. Note that the MAE facet does not include the group ($\delta \tau$) because $\delta$ (the performance measure used) is no longer an uncertain parameter in this setting.}
\label{fig:SA}
\end{figure}

When the aim is to rank the model inputs, the selection of the performance measure ($\delta$) has the highest first-order effect in the accuracy of the estimators (Fig.~\ref{fig:SA}a). The parameter $\delta$ is responsible for between 20\% (Azzini and Rosati) and 30\% (Glen and Isaacs) of the variance in the final $r$ value. On average, all estimators perform better when the rank is conducted on Savage scores ($\delta=2$), i.e. when the focus is on ranking the most important model inputs only (Figs.~S8--S15). As for the distribution of the model inputs ($\phi$), it has a first-order effect in the accuracy of Azzini and Rosati ($\approx10$\%), Jansen and  Janon / Monod ($\approx15$\%) and Razavi and Gupta ($\approx20$\%) regardless of whether the aim is a factor prioritization ($r$) or approaching the ``true'' indices (MAE). The performance of these estimators drops perceptibly when the model inputs are distributed as $Beta(8,2)$ or $Beta(2,8)$ ($\phi=3$ and $\phi=4$, Figs. S8-S23), suggesting that they may be especially stressed by skewed distributions. The selection of random or quasi-random numbers during the construction of the sample matrix ($\tau$) also directly conditions the accuracy of several estimators. If the aim is to approach the ``true'' indices (MAE), $\tau$ conveys from 17\% (Azzini and Rosati) to $\approx30$\% (Glen and Isaacs) of the model output variance, with all estimators except Razavi and Gupta performing better on quasi-random numbers ($\tau=2$, Figs.~S16--S23). In a factor prioritization setting, $\tau$ is mostly influential through interactions. Interestingly, the proportion of active second and third-order interactions ($k_2,k_3$) does not alter the performance of any estimator in any of the settings examined.

To better understand the structure of the sensitivities, we compute Sobol' indices after grouping individual parameters in three clusters, which we define based on their commonalities: the first group includes $(\delta,\tau)$ and reflects the influence of those parameters that can be defined by the sensitivity analyst during the setting of the benchmark exercise. The second combines ($\varepsilon,k_2,k_3,\phi$) and examines the overall impact of the model functional form, referred to as $f(x)$, which is often beyond the analyst's grasp. Finally, the third group includes $(N_t,k)$ only and assesses the influence of the sampling design in the accuracy of the estimators (we assume that the total number of model runs, besides being conditioned by the computing resources at hand, is also partially determined by the joint effect of the model dimensionality and the use of either a $\bm{B}$, $\bm{A}_{B} ^{(i)})$, $\bm{B}_{A} ^{(i)}$ or $\bm{C}_{B} ^{(i)}$ matrices) (Fig~\ref{fig:SA}b).

The uncertainty in the functional form of the model [$f(x)$] is responsible for approximately 20\% of the variance in the performance of Azzini and Rosati, Janon/Monod or Jansen in a factor prioritization setting. For Glen and Isaacs, Homma and Saltelli, pseudo-Owen or Saltelli, $f(x)$ is influential only through interactions with the other clusters. When the MAE is the performance measure of interest, $f(x)$ has a much stronger influence in the accuracy of the estimators than the couple $(N_t,k)$, especially in the case of Glen and Isaacs ($\approx 40$\%). In any case, the accuracy of the estimators is significantly conditioned by interactions between the benchmark parameters. The sum of all individual $S_i$ indices plus the $S_i$ index of the $(N_t,k)$ cluster only explains from $\approx 45$\% (Saltelli) to $\approx 70$\% (Glen and Isaacs) of the estimators' variance in ranking the model inputs, and from $\approx 24$\% (pseudo-Owen) to $\approx 60$\% (Razavi and Gupta) of the variance in approaching the ``true'' indices.

\section{Discussion and conclusions}
\label{sec:discussion}
Here we design an eight-dimension background for variance-based total-order estimators to confront and prove their value in an unparalleled range of SA scenarios. By randomizing the parameters that condition their performance, we obtain a comprehensive picture of the advantages and disadvantages of each estimator and identify which particular benchmark factors make them more prone to error. Our work thus provides a thorough empirical assessment of state-of-the-art total-order estimators and contributes to define best practices in variance-based SA. The study also aligns with previous works focused on testing the robustness of the tools available to sensitivity analysts, a line of inquiry that can be described as a \emph{sensitivity analysis of a sensitivity analysis} (SA of SA) \parencite{Puy2020}. 

Our results provide support to the assumption that the scope of previous benchmark studies is limited by the plethora of non-unique choices taken during the setting of the analysis \parencite{Becker2020}. We have observed that almost all decisions have a non-negligible effect: from the selection of the sampling method to the choice of the performance measure, the design prioritized by the analyst can influence the performance of the estimator in a non-obvious way, namely through interactions. The importance of non-additivities in conditioning performance suggests that the benchmark of sensitivity estimators should no longer rely on statistical designs that change one parameter at a time (usually the number of model runs and, more rarely, the test function \cite{Janon2014, Azzini2019, Azzini2020, Saltelli2010a, LoPiano2021, Razavi2016a, Razavi2016b, Owen2013, Puy2020}). Such setting reduces the uncertain space to a minimum and misses the effects that the interactions between the benchmark parameters have in the final accuracy of the estimator. If global SA is the recommended practice to fully explore the uncertainty space of models, sensitivity estimators, being algorithms themselves, should be likewise validated \parencite{Puy2020}. 

Our approach also compensates the lack of studies on the theoretical properties of estimators in the sensitivity analysis literature (see for instance \cite{Azzini2020a, Jansen1999}), and allows a more detailed examination of their performance than theoretical comparisons. Empirical studies like ours mirror the numerical character of sensitivity analysis when the indices can not be analytically calculated, which is most of the time in ``real-world'' mathematical modeling.

Two recommendations emerge from our work: the estimators by Razavi and Gupta, Jansen, Janon / Monod or Azzini and Rosati should be preferred when the aim is to rank the model inputs. Jansen, Janon/Monod or Azzini and Rosati should also be prioritized if the goal is to estimate the ``true'' total-order indices. The drop in performance of Razavi and Gupta in the second setting may be explained by a bias at a lower sample sizes, i.e. a consistent over-estimation of all total-order indices. This is because their estimator relies on a constant mean assumption whose validity degrades with larger values of $\Delta h$ \cite{Razavi2016a, Razavi2016b}. In order to remove this bias, $\Delta h$ should take very small values (e.g., $\Delta h = 0.01$), which may not be computationally feasible. Since the direction of this bias is the same for all parameters it only affects the calculation of the ``true'' total-order indices, not the capacity of the estimator to properly rank the model inputs.

It is also worth stating that Razavi and Gupta is the only estimator studied here that require the analyst to define a tuning parameter, $\Delta h$. In this paper we have set $\Delta h=0.2$ after some preliminary trials with the estimator; other works have used different values (e.g. $\Delta h = 0.002$, $\Delta h = 0.1$, $\Delta h = 0.3$; \cite{Becker2020, Razavi2016a, Razavi2016b}). Selecting the most appropriate value for a given tuning parameter is not an obvious choice and this uncertainty can make an estimator volatile, as shown by \textcite{Puy2020} in the case of the PAWN index.

The fact that Glen and Isaacs, Homma and Saltelli, Saltelli and pseudo-Owen do not perform as well in properly ranking the model inputs and approaching the ``true'' total-order indices may be partially explained by their less efficient computation of elementary effects: by allowing the production of negative terms in the numerator these estimators also permit the production of negative total-order indices, thus leading to biased rankings or sensitivity indices. In the case of Saltelli, the use of a $\bm{B}$ matrix at the numerator and an $\bm{A}$ matrix at the denominator exacerbates its volatility (Table~\ref{tab:Ti_estimators}, Nº 5). Such inconsistency was corrected in \textcite{Saltelli2010a}.

The consistent robustness of Jansen, Janon/Monod and Azzini and Rosati makes their sensitivity to the uncertain parameters studied here almost negligible. They are already highly optimized estimators with not much room for improvement. Most of their performance is conditioned by the first and total-order effects of the model form jointly with the underlying probability distributions ($f(x)$ in Fig.~\ref{fig:SA}b), as well as by their sampling design ($N_t,k$), which are in any case beyond the analyst's control. As for the rest, their accuracy might be enhanced by allocating a larger number of model runs per input (if computationally affordable), and especially in the case of Homma and Saltelli, Saltelli and Glen and Isaacs, by restricting their use to low-dimensional models ($k<10$) and sensitivity settings that only require ranking the most important parameters (a \emph{restricted} factor prioritisation setting; \cite{Saltelli2008}). Nevertheless, their substantial volatility is considerably driven by non-additivities, a combination that makes them hard to tame and should raise caution about their use in any modeling exercise.

Our results slightly differ from \textcite{Becker2020}'s, who observed that Jansen outperformed Janon/Monod under a factor prioritization setting. We did not find any significant difference between these estimators. Although our metafunction approach is based on \textcite{Becker2020}'s, our study tests the accuracy of estimators in a larger uncertain space as we also account for the stress introduced by changes in the sampling method $\tau$, the underlying probability distributions $\phi$ or the performance measure selected $\delta$. These differences may account for the slightly different results obtained between the two papers.

Our analysis can be extended to other sensitivity estimators (i.e. moment-independent like entropy-based \cite{Liu2006a}; the $\delta$-measure \cite{Borgonovo2007}; or the PAWN index, \cite{Pianosi2015, Pianosi2018}). Moreover, it holds potential to be used overall as a standard crash test every time a new sensitivity estimator is introduced to the modeling community. One of its advantages is its flexibility: \textcite{Becker2020}'s metafunction can be easily extended with new univariate functions or probability distributions, and the settings modified to check performance under different degrees of non-additivities or in a larger $(N_t,k)$ space. With some slight modifications it should also allow to produce functions with dominant low-order or high-order terms, labeled as Type B and C by \textcite{Kucherenko2011}. This should prompt developers of sensitivity indices to severely stress their estimators so the modeling community and decision-makers fully appraise how they deal with uncertainties.

\section{Code availability}
The \texttt{R} code to replicate our results is available in \textcite{Puy2020c} and in GitHub (\url{https://github.com/arnaldpuy/battle_estimators}). The uncertainty and sensitivity analysis have been carried out with the \texttt{R} package \texttt{sensobol} \cite{Puyk}, which also includes the test function used in this study.

\section{Acknowledgements}
We thank Saman Razavi for his insights on the behavior of the Razavi and Gupta estimator. This work has been funded by the European Commission (Marie Sk\l{}odowska-Curie Global Fellowship, grant number 792178 to A.P.). 

\printbibliography

\end{document}


\pagenumbering{arabic}
\date{}
\title{A comprehensive comparison of total-order estimators for global sensitivity analysis \\ \vspace{3mm} \large{Supplementary Materials}}
\author[1,2]{Arnald Puy\thanks{Corresponding author}}
\author[3]{William Becker}
\author[4]{Samuele Lo Piano}
\author[2,3]{Andrea Saltelli}

\affil[1]{\footnotesize{\textit{Department of Ecology and Evolutionary Biology, M31 Guyot Hall, Princeton University, New Jersey 08544, USA. E-Mail: apuy@princeton.edu}}}

\affil[2]{\footnotesize{\textit{Centre for the Study of the Sciences and the Humanities (SVT), University of Bergen, Parkveien 9, PB 7805, 5020 Bergen, Norway.}}}

\affil[3]{\footnotesize{\textit{European Commission, Joint Research Centre, Via Enrico Fermi, 2749, 21027 Ispra VA, Italy}}}

\affil[4]{\footnotesize{\textit{University of Reading, School of the Built Environment, JJ Thompson Building, Whiteknights Campus, Reading, RG6 6AF, United Kingdom}}}

\maketitle

\tableofcontents

\newpage

\beginsupplement

\section{Razavi and Gupta's estimator (VARS)}
Unlike the other total-order estimators examined in our paper, Razavi and Gupta's VARS (for Variogram Analysis of Response Surfaces \cite{Razavi2016a, Razavi2016b}) relies on the variogram $\gamma(.)$ and covariogram $C(.)$ functions  to compute what they call the VARS-TO, for VARS Total-Order index. 

Let us consider a function of factors $\bm{x}=(x_1, x_2, ..., x_k)\in \mathbb{R}^k$. If $\bm{x}_A$ and $\bm{x}_B$ are two generic points separated by a distance $\bm{h}$, then the variogram is calculated as

\begin{equation}
\gamma(\bm{x}_A-\bm{x}_B) = \frac{1}{2}V \left [y(\bm{x}_A) - y(\bm{x}_B) \right ]
\end{equation}

and the covariogram as

\begin{equation}
C(\bm{x}_A-\bm{x}_B) = COV \left [y(\bm{x}_A),  y(\bm{x}_B) \right ]
\end{equation}

Note that

\begin{equation}
V \left [y(\bm{x}_A) - y(\bm{x}_B) \right ] = V \left [y(\bm{x}_A) \right ] + V \left [y(\bm{x}_B) \right ] - 2COV \left [ y(\bm{x}_A), y(\bm{x}_B) \right ]
\end{equation}

and since $V \left [ y(\bm{x}_A) \right ] = V \left [ y(\bm{x}_B) \right ]$, then 

\begin{equation}
\gamma (\bm{x}_A - \bm{x}_B) = V \left [ y(\bm{x}) \right ] - C(\bm{x}_A, \bm{x}_B)
\label{eq:variogram}
\end{equation}

In order to obtain the total-order effect $T_i$, the variogram and covariogram are computed on all couples of points spaced $h_i$ along the $x_i$ axis, with all other factors being kept fixed. Thus equation~\ref{eq:variogram} becomes 

\begin{equation}
\gamma_{x^*_\sim i}(h_i)=V(y|x^*_\sim i)-C_{x^*_\sim i}(h_i)
\label{eq:var_fixed}
\end{equation}

where $x^*_\sim i$ is a fixed point in the space of non-$x_i$. \textcite{Razavi2016a, Razavi2016b} suggest to take the mean value across the factors' space on both sides of equation~\ref{eq:var_fixed}, thus obtaining

\begin{equation}
E_{x^*_\sim{i}} \left [ \gamma_{x^*_\sim i}(h_i) \right ]=E_{x^*_\sim{i}}  \left [ V(y|x^*_{\sim i}) \right ] -E_{x^*_\sim{i}}  \left [ C_{x^*_\sim i}(h_i) \right ]
\end{equation}

which can also be written as

\begin{equation}
E_{x^*_\sim{i}} \left [ \gamma_{x^*_\sim i}(h_i) \right ]=V(y)T_i -E_{x^*_\sim{i}}  \left [ C_{x^*_\sim i}(h_i) \right ]
\end{equation}

and therefore

\begin{equation}
T_i=\frac{E_{x^*_\sim i}\left [ \gamma_{x^*_\sim i}(h_i)\right] + E_{x^*_\sim i} \left [ C_{x^*_\sim i}(h_i) \right ] }{V(y)}
\label{eq:SM_VARS_ti}
\end{equation}

The sampling scheme for VARS does not rely on $\textbf{A}, \textbf{B}, \textbf{A}_{B}^{(i)}...$ matrices, but on star centers and cross sections. Star centers are $N$ random points sampled across the input space. For each of these stars, $k$ cross sections of points spaced $\Delta h$ apart are generated, including and passing through the star center. Overall, the computational cost of VARS amounts to $N_t=N\left [k((1 /\Delta h )-1) + 1 \right]$.

\newpage

\section{Figures}

\begin{figure}[ht]
\centering
\includegraphics[keepaspectratio]{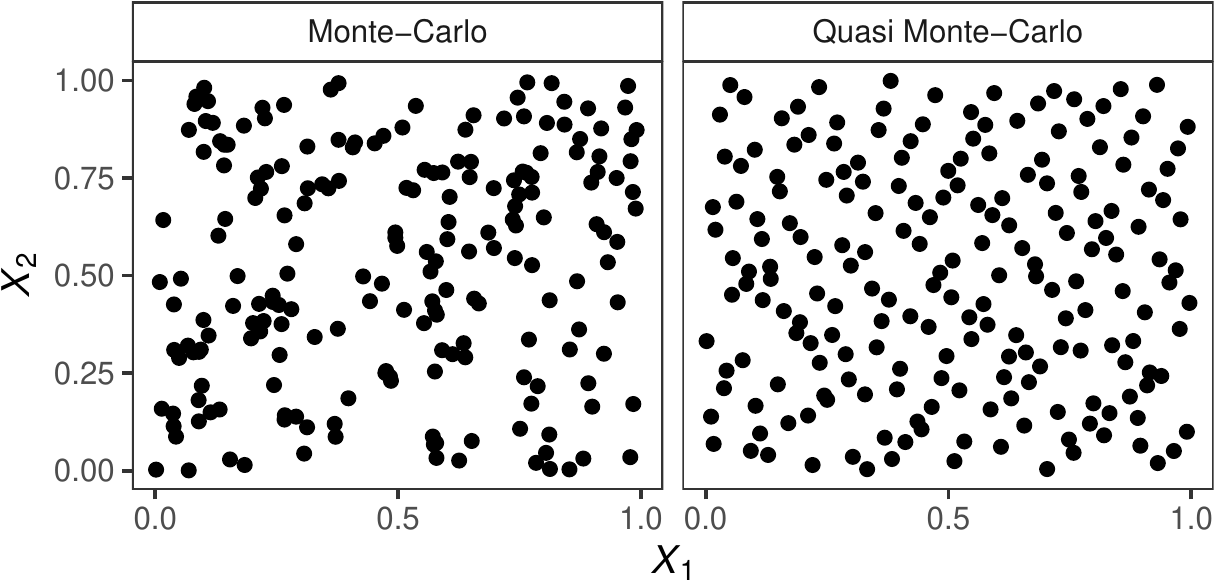}
\caption{Examples of Monte-Carlo and Quasi Monte-Carlo sampling in two dimensions. $N=200$.}
\label{fig:SM_sampling_method}
\end{figure}

\begin{figure}[ht]
\centering
\includegraphics[keepaspectratio]{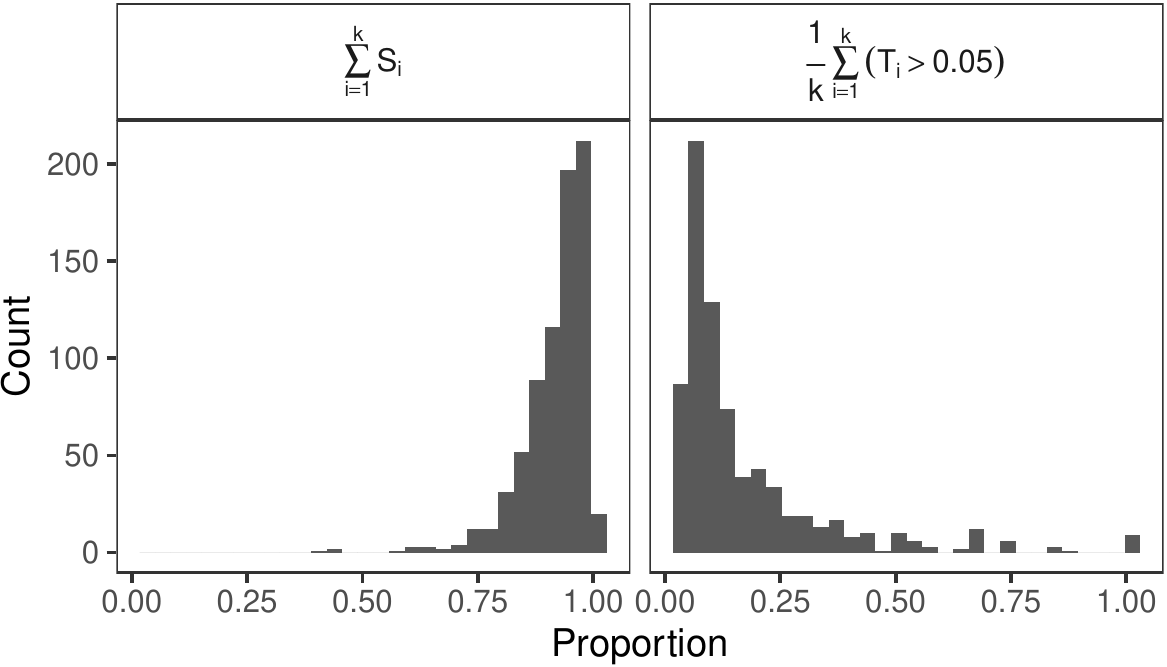}
\caption{Proportion of the total sum of first-order effects and of the active model inputs (defined as $T_i>0.05$) after 1000 random metafunction calls with $k\in(3,100)$. Note how the sum of first-order effects clusters around $0.8$ (thus evidencing the production of non-additivities) and how, on average, the number of active model inputs revolves around 10--20\%, thus reproducing the Pareto principle.}
\label{fig:SM_prove_meta}
\end{figure}

\begin{figure}[ht]
\centering
\includegraphics[keepaspectratio]{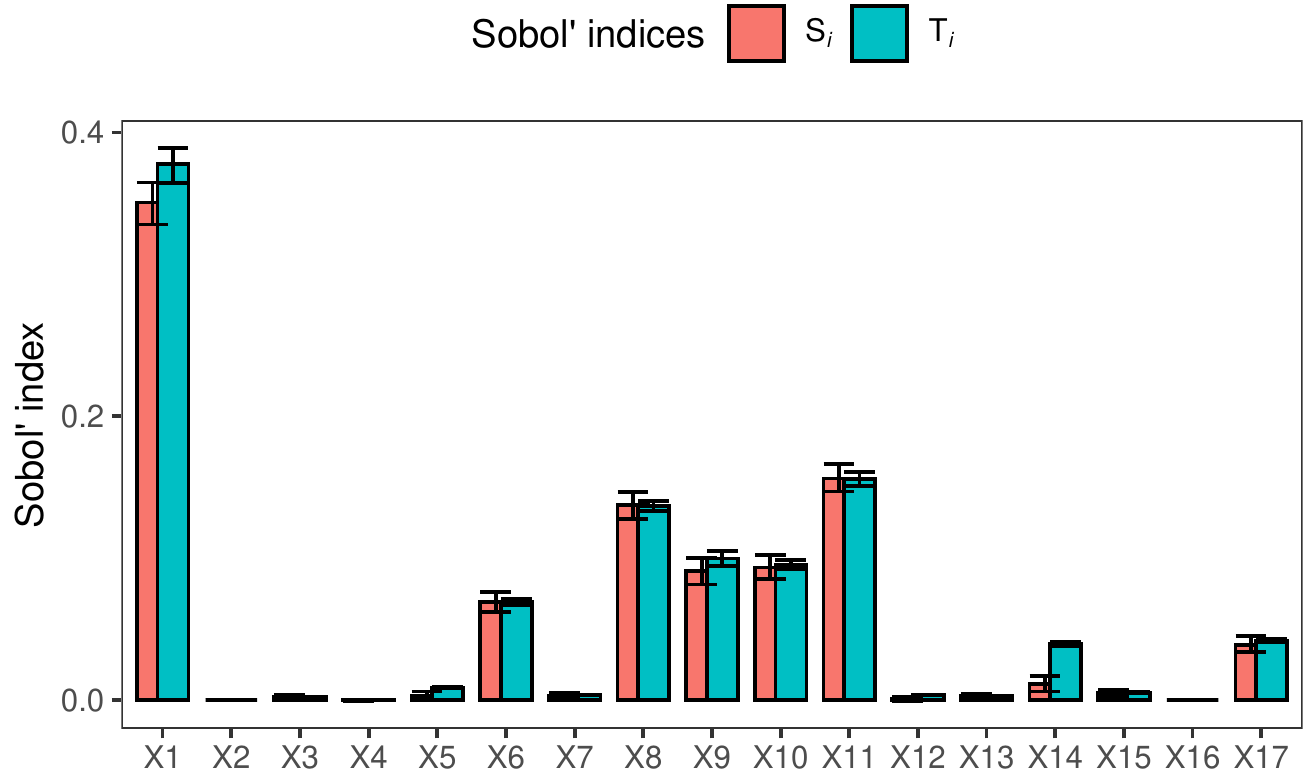}
\caption{Sobol' $T_i$ indices obtained after a run of the metafunction with the following parameter settings: $N=10^4$, $k=17$, $k_2=0.5$, $k_3=0.2$, $\varepsilon=666$. The error bars reflect the 95\% confidence intervals after bootstrapping ($R=10^2$). The indices have been computed with the \textcite{Jansen1999} estimator.}
\label{fig:SM_metafunction}
\end{figure}

\begin{figure}
\centering
\includegraphics[keepaspectratio]{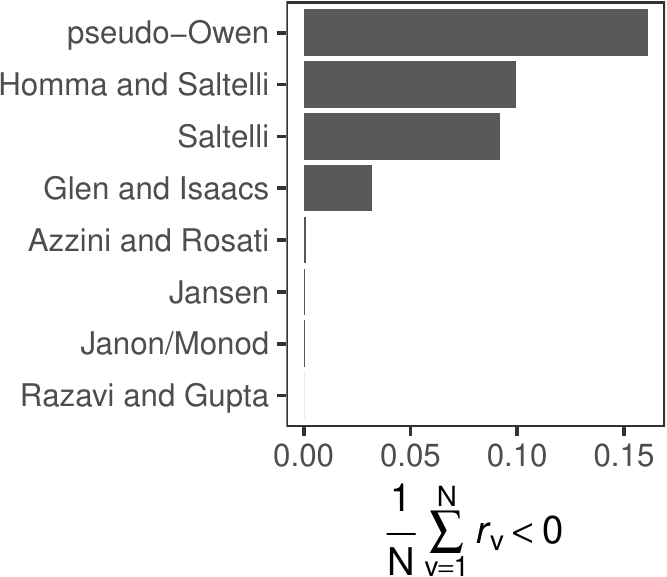}
\caption{Proportion of model runs yielding $r<0$.}
\label{fig:SM_negative}
\end{figure}

\begin{figure}
\centering
\includegraphics[keepaspectratio]{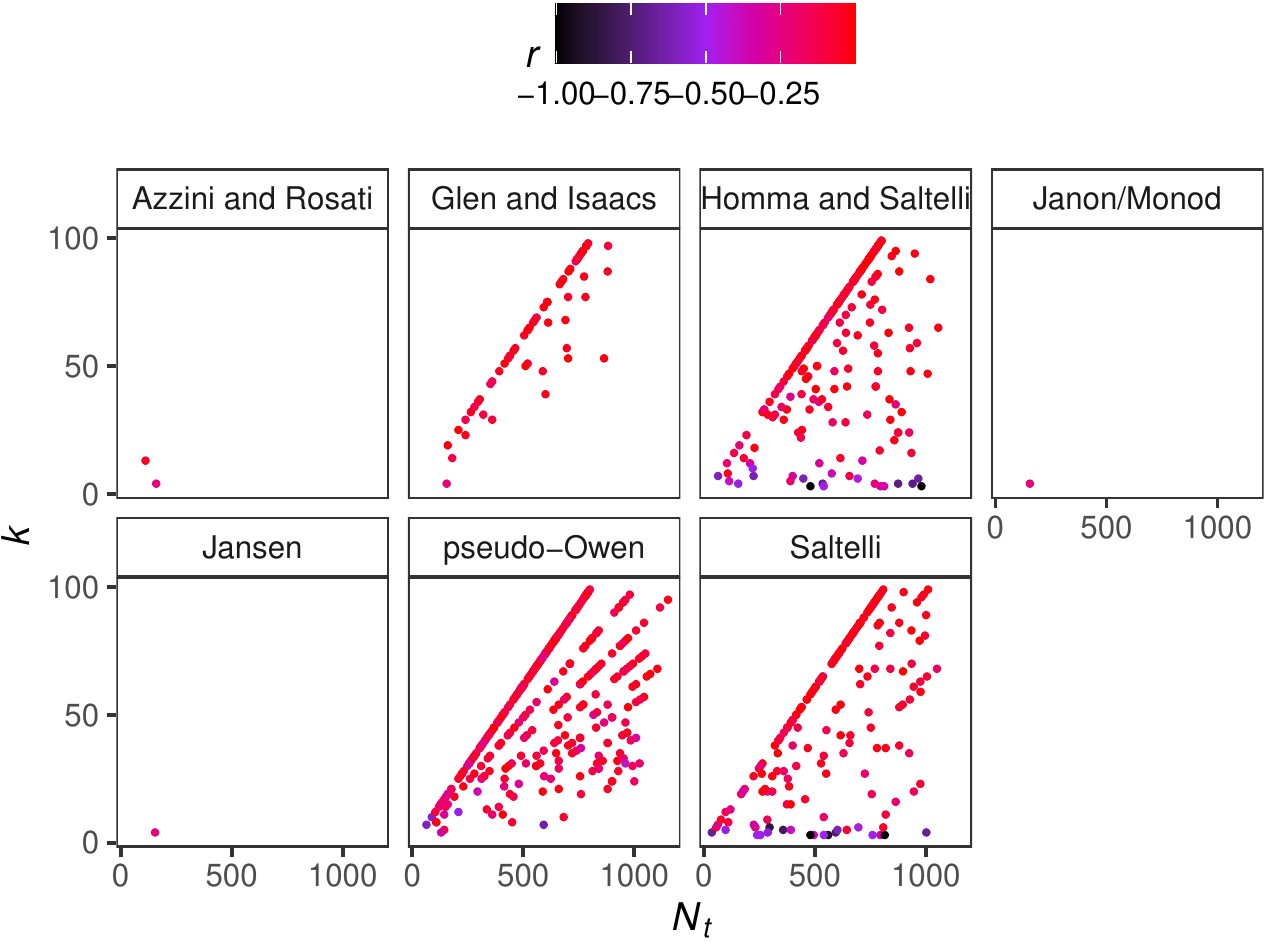}
\caption{Scatter of the total number of model runs $N_t$ against the function dimensionality $k$ only for $r<0$.}
\label{fig:SM_negative_map}
\end{figure}

\begin{figure}
\centering
\includegraphics[keepaspectratio]{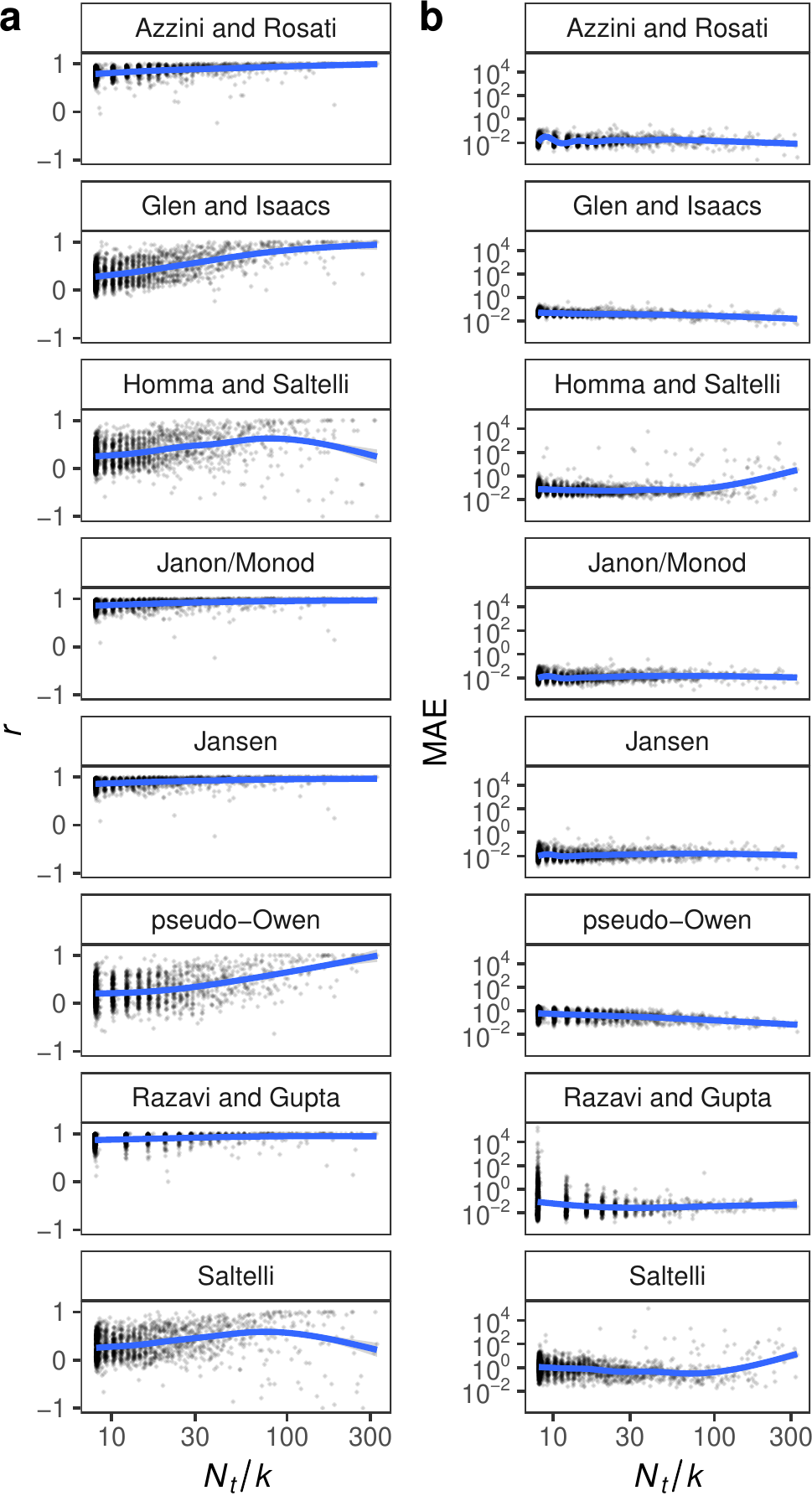}
\caption{Scatterplot of the correlation between $\bm{T}_i$ and $\bm{\hat{T}}_i$ ($r$) against the number of model runs allocated per model input $(Nt / k)$.}
\label{fig:SM_ratio}
\end{figure}

\begin{figure}
\centering
\includegraphics[keepaspectratio]{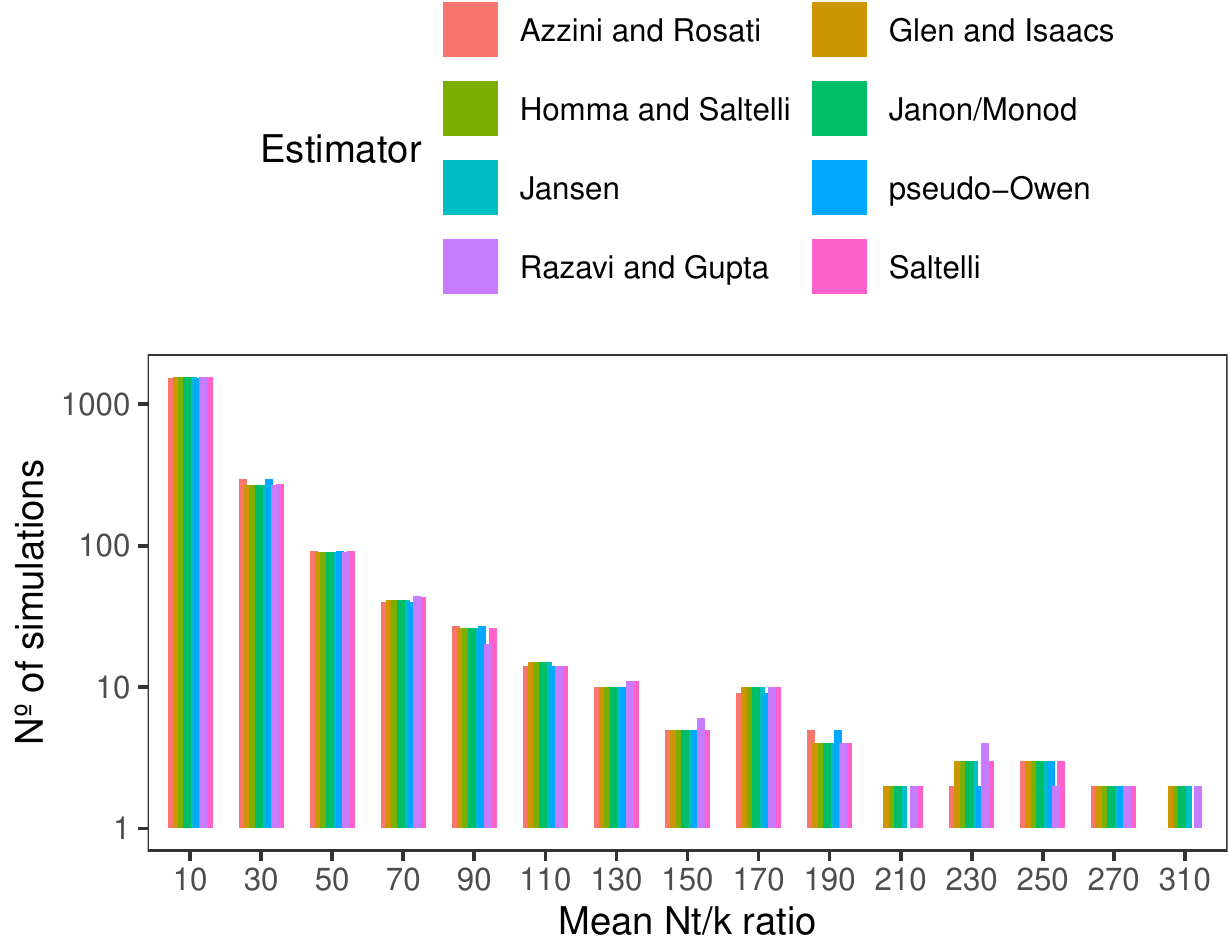}
\caption{Bar plot with the number of simulations conducted in each of the $N_t / k$ comparments assessed. All estimators have approximately the same number of simulations in each compartment.}
\label{fig:SM_ntk_ratio}
\end{figure}

\foreach \x in {1,...,16} { 
\begin{figure}
\centering
\includegraphics[keepaspectratio, height=\textheight, width=\textwidth]{scatterplots_sens-\x}
\caption{Scatterplots of the model inputs against the model output. The red dots show the mean value in each bin (we have set the number of bins arbitrarily at 30).}
\label{fig:SM_scatters}
\end{figure}
}

\clearpage

\clearpage

\printbibliography